\documentclass[prd,superscriptaddress,amsfonts,amssymb,amsmath,showpacs,twocolumn]{revtex4-2}

\usepackage{bm}
\usepackage{amsfonts}
\usepackage{latexsym}
\usepackage[latin1]{inputenc}
\usepackage{graphicx}
\usepackage{amsmath}
\usepackage{palatino}
\usepackage{mathpazo}
\usepackage{textcomp}
\usepackage{float}
\usepackage{booktabs}
\usepackage{dcolumn}
\usepackage{ragged2e}
\usepackage{hyperref}
\hypersetup{colorlinks,citecolor=blue}
\hypersetup{colorlinks=true,linkcolor=red,filecolor=magenta,    urlcolor=blue}
\usepackage{amsmath}
\usepackage{graphicx}
\usepackage{makecell}
\usepackage{hyperref}
\usepackage[mathlines]{lineno}
\usepackage{amssymb}
\usepackage{enumitem}
\usepackage{mathtools}
\usepackage{mathrsfs}
\usepackage{setspace}
\usepackage{color}
\usepackage{graphicx}
\usepackage[dvipsnames]{xcolor}
\usepackage{mathalfa}
\usepackage{calligra}
\usepackage[misc]{ifsym}
\DeclareMathAlphabet{\mathpzc}{OT1}{pzc}{m}{it}
\DeclareMathAlphabet{\mathcalligra}{T1}{calligra}{m}{n}
\hypersetup{colorlinks=true,citecolor=blue,linkcolor=red,urlcolor=magenta}
\usepackage{xcolor}
\usepackage{orcidlink}
\usepackage[caption=false]{subfig}
\usepackage{commath}
\def\jnl@style{}
\def\aaref@jnl#1{{\jnl@style#1}}

\def\aaref@jnl#1{{\jnl@style#1}}

\def\aj{\aaref@jnl{AJ}}                   
\def\apj{\aaref@jnl{ApJ}}                 
\def\apjl{\aaref@jnl{ApJ}}                
\def\apjs{\aaref@jnl{ApJS}}               
\def\apss{\aaref@jnl{Ap\&SS}}             
\def\aap{\aaref@jnl{A\&A}}                
\def\aapr{\aaref@jnl{A\&A~Rev.}}          
\def\aaps{\aaref@jnl{A\&AS}}              
\def\mnras{\aaref@jnl{Mon.~Not.~Roy.~Astron.~Soc.}}             
\def\prd{\aaref@jnl{Phys.~Rev.~D}}        
\def\plb{\aaref@jnl{Phys.~Lett.~B}}        
\def\prc{\aaref@jnl{Phys.~Rev.~C}}  
\def\prl{\aaref@jnl{Phys.~Rev.~Lett.}}    
\def\qjras{\aaref@jnl{QJRAS}}             
\def\skytel{\aaref@jnl{S\&T}}             
\def\ssr{\aaref@jnl{Space~Sci.~Rev.}}     
\def\zap{\aaref@jnl{ZAp}}                 
\def\nat{\aaref@jnl{Nature}}              
\def\aplett{\aaref@jnl{Astrophys.~Lett.}} 
\def\apspr{\aaref@jnl{Astrophys.~Space~Phys.~Res.}} 
\def\physrep{\aaref@jnl{Phys.~Rep.}}      
\def\physscr{\aaref@jnl{Phys.~Scr}}       
\def\commat{\aaref@jnl{Comm.~Math.~Phys.}}              
\def\science{\aaref@jnl{Science}}               
\def\cqg{\aaref@jnl{Classical Quant.~Grav.}}            
\def\jpcs{\aaref@jnl{JPCS}}                                     
\def\ijmpd{\aaref@jnl{Int.~J.~Mod.~Phys.~D}}                    
\def\grg{\aaref@jnl{Gen.~Relat.~Gravit.}}               
\def\rpp{\aaref@jnl{Rep.~Prog.~Phys.}}          
\def\npa{\aaref@jnl{Nucl.~Phys.~A}}        
\def\lrr{\aaref@jnl{Living Rev.~Rel.}}                   
\def\jcap{\aaref@jnl{J.~Cosmology Astropart.~Phys.}}    
\def\rmp{\aaref@jnl{Rev.~Mod.~Phys.}}   
\def\epjc{\aaref@jnl{Eur.~Phys.~J.~C}}

\allowdisplaybreaks[1]
\renewcommand{\arraystretch}{1.1}
\begin{document}

\preprint{APS/123-QED}

\title{Exploring wormhole solutions in curvature-matter coupling gravity supported by noncommutative geometry and conformal symmetry}

\author{N. S. Kavya\orcidlink{0000-0001-8561-130X}}
\email{kavya.samak.10@gmail.com}
\affiliation{Department of P.G. Studies and Research in Mathematics,
 \\
 Kuvempu University, Shankaraghatta, Shivamogga 577451, Karnataka, INDIA
}%

\author{G. Mustafa\orcidlink{0000-0003-1409-2009}}%
\email{gmustafa3828@gmail.com (Corresponding Author)}
\affiliation{
 Department of Physics, Zhejiang Normal University, Jinhua, 321004, People's Republic of China.
}%

\author{V. Venkatesha\orcidlink{0000-0002-2799-2535}}%
 \email{vensmath@gmail.com}
\affiliation{Department of P.G. Studies and Research in Mathematics,
 \\
 Kuvempu University, Shankaraghatta, Shivamogga 577451, Karnataka, INDIA
}%

\author{P.K. Sahoo\orcidlink{0000-0003-2130-8832}}
\email{pksahoo@hyderabad.bits-pilani.ac.in}
\affiliation{
 Department of Mathematics, Birla Institute of Technology and Science-Pilani,\\
 Hyderabad Campus, Hyderabad 500078, INDIA
}%



\date{\today}

\begin{abstract}
This article explores new physically viable wormhole solutions within the framework of $\mathpzc{f}(\mathcal{R},\mathscr{L}_m)$ gravity theory, incorporating noncommutative backgrounds and conformal symmetries. The study investigates the impact of model parameters on the existence and properties of wormholes. The derived shape function is found to obey all the required criteria. Specific attention is given to traceless wormholes with Gaussian and Lorentzian distributions, investigating the behavior of the shape functions and energy conditions. In both cases, the presence of exotic fluid is confirmed. 
\begin{description}
\item[Keywords]
Traversable wormhole, $\mathpzc{f}(\mathcal{R},\mathscr{L}_m)$ gravity, energy conditions, noncommutative geometry,\\ conformal motion

\end{description}
\end{abstract}

\maketitle

\tableofcontents
\section{Introduction}\label{I}
    Wormholes have gained significant attention in recent years due to their potential to connect different points in spacetime topology. The concept of traversable wormholes was introduced by Morris and Thorne \cite{morrisandthorne}, who explored the possibility of constructing wormholes that could be matter-traversable and even facilitate time travel. The existence of such a topological entity has been the subject of extensive exploration and theoretical investigation. In the context of classical general relativity, these wormholes exhibit the presence of exotic matter, for which the energy-momentum tensor (EMT) violates the null energy condition (NEC). However, studies in modified gravity reveal that incorporating higher-order curvature terms allows the stress-energy tensor of ordinary matter to satisfy energy conditions while still supporting the exotic geometries of wormholes \cite{ec1,ec2,ec3,ec4}. Furthermore, the literature encompasses various other types of wormholes where the traversing path does not necessitate the presence of exotic matter \cite{exotic1,exotic2,exotic3,exotic4}. To examine wormhole solutions in the context of modified theories, numerous works have been carried out (see  \cite{ref1,ref2,ref3,ref4,ref5,ref6,ref7,ref8,ref9,ref10,ref11} and the references therein).

    One of the interesting approaches in the exploration of the manifold structure is through modifications of the matter source. This is facilitated by the concept of noncommutative geometries. It offers an effective framework to address both spacetime geometry deformations and quantization processes \cite{ncg,ncg1,ncg2,nc1,nc7,nc8,nc9,nc10,nc11,nc,pkfk1,pkfk2,intrinsic,lorentzianWH,ncref}. In the context of D-brane \cite{nc2}, the coordinates of spacetime are treated as noncommutative operators satisfying the relation $[y^a,y^b]=\mathit{i}\Theta^{ab}$, where $\Theta^{ab}$ is a second-order antisymmetric matrix \cite{nc3,nc4,nc5,nc6}. This formulation indicates the discretization of spacetime, replacing point-like structures with smeared objects. The smearing effect is achieved by substituting the Dirac delta function with Gaussian and Lorentzian distributions characterized by a minimal length scale $\sqrt{\Theta}$.
    
    For the static spherically symmetric point-like gravitational source with total mass $M$, Gaussian and Lorentzian distribution of energy densities are represented by \cite{intrinsic,nc,jim1,jim2},
    
    \begin{gather}
        \label{gauss}\rho= \frac{M e^{-\frac{r^2}{4 \Theta }}}{8 \pi ^{\frac{3}{2}} \Theta ^{\frac{3}{2}}},
        \end{gather}
    and
      \begin{gather}
        \label{lorentz}\rho=\frac{\sqrt{\Theta } M}{\pi ^2 \left(\Theta +r^2\right)^2}.
    \end{gather}

    These choices support the concept that the source exhibits a distributed or smeared nature instead of being concentrated at a single point. This characteristic arises due to the inherent uncertainty associated with the coordinate commutator. Moreover, the impact of noncommutativity becomes prominent in the domain having the origin, specifically when the radial distance is smaller than the scale parameter $\Theta$. Within this local region, the effects of noncommutativity act as a regularizer for both the radial and tangential pressures, as well as the matter density. By employing the specific choices given by equations \eqref{gauss} and \eqref{lorentz}, the physical parameters, particularly the energy density, remain finite and asymptotically approach zero as one moves towards points that are far away from the origin. This behavior supports the existence of a vacuum solution in distant regions.

    In our study to explore a profound connection between geometry and matter with governing equations, the employment of inheritance symmetry proves to be a valuable technique \cite{ckv1,ckv2,ckv3,ckv4,ckv5,ckv6}. In particular, the notable inheritance symmetry known as conformal killing vectors (CKVs) emerges as a significant avenue of exploration. CKVs exhibit the property of conformal invariance, wherein the metric can be conformally mapped onto itself along the vector $\eta$ through the action of the Lie derivative operator denoted as \cite{ckv1},

    \begin{equation}\label{eq:ckv}
        \mathcal{L}_{\eta}g_{ab}=g_{na}\eta^{n}_{;b}+g_{an}\eta^{n}_{;b}=\Psi(r) g_{ab},
    \end{equation}
    with $\Psi$, $\eta^n$, and $g_{ab}$ representing the conformal factor, CVKs, and metric tensor respectively. 

    The recent discovery of the cosmological aspects, initially supported by observations in \cite{acc}, has prompted the need to uncover a compelling and trustworthy explanation for the significant phenomena. This also raised the issue of the Universe's peculiar composition. Therefore, it becomes imperative to explore broader modifications to standard general relativity. A prominent interest lies in generalized gravity models that incorporate connections between the geometry of spacetime and the presence of matter. 
    
    In the literature, various geometry theories, including $\mathpzc{f}(\mathcal{R})$, $\mathpzc{f}(\mathcal{G})$, $\mathpzc{f}(\mathcal{Q})$, and more, have been explored. Numerous studies have investigated these theories to understand their implications in the realms of astronomy and cosmology. These investigations have shed light on a range of important phenomena, such as the late-time cosmic acceleration, the exclusion of specific dark matter candidates through the analysis of massive test particles, and the integration of inflationary models with dark energy. Ref. \cite{fr} provides reasonable explanations for these phenomena within the framework of $\mathpzc{f}(\mathcal{R})$ theories. 

    The geometry theories modify the gravitational Lagrangian by considering arbitrary functions of geometric elements. An extension of this approach involves incorporating the matter Lagrangian along with geometric description, resulting in the coupling of the geometry and matter content of the universe. Lobo et al., in \cite{coupling} have presented several physical aspects of such curvature-matter coupling theories.
    
    In the present letter, we study wormhole solutions exhibiting conformal motion with Gaussian and Lorentzian noncommutative backgrounds. The examination of the solution is carried out in the context of curvature-matter coupling gravity. An overview of $\mathpzc{f}({\mathcal{R},\mathscr{L}_m})$ theory and its theoretical and physical motivations are given in \sectionautorefname~\ref{IIa}. There are primarily three factors in our current analysis of wormhole solutions. Firstly, the choice of coupling theory, which is constructed using the standard Levi-Civita connection and includes an arbitrary function of matter Lagrangian, in addition to the Ricci scalar. As an extension of the f(R) theory, this theory can address numerous cosmological aspects. Secondly, the implementation of noncommutative geometry, which can provide quantization effects through the discretization of spacetime. Lastly, we consider conformal symmetry. These transformations preserve angles in spacetime and inherit symmetry. The primary objective of our current work is to examine traversable wormhole solutions within the framework of these modifications.

    Article structure: \sectionautorefname~ \ref{II} provides a detailed presentation of the mathematical formulation of the modified theory, including the governing equations, the exploration of wormhole solutions within $\mathpzc{f}(\mathcal{R},\mathscr{L}_m)$ gravity, and an analysis of the energy conditions. Moving on to \sectionautorefname~\ref{III}, we delve into the examination of the wormhole model featuring Gaussian and Lorentzian distributions. Within this section, we derive the corresponding shape functions and investigate the influence of model parameters on these functions as well as the energy conditions. The geometry of wormholes involving traceless fluids is carefully assessed in \sectionautorefname~\ref{IV}. Lastly, \sectionautorefname~\ref{V} offers a comprehensive discussion of the findings obtained throughout the study.
	
\section{Mathematical Formulation of the Modified Theory}\label{II}
\subsection{Overview of $\mathpzc{f}(\mathcal{R},\mathscr{L}_m)$ gravity}\label{IIa}

Observational constraints have highlighted certain limitations of GR at different scales, such as the quantum and galactic scales. These shortcomings necessitate modifications to the standard action in order to preserve GR as the fundamental theory of gravity. Among these, a prominent one is $\mathpzc{f}(\mathcal{R})$ modified geometry theory, which can fairly accommodate observational data. Interestingly, $\mathpzc{f}(\mathcal{R},\mathscr{L}_m)$ gravity, a generalized version of $\mathpzc{f}(\mathcal{R})$ theory, presented in \cite{frlm}, incorporates modifications in both the geometry and matter sectors. The explicit coupling between geometry and the matter results in a non-vanishing covariant derivative of the Energy-Momentum Tensor (EMT), leading to deviations from geodesic motion and violation of the equivalence principle. Different forms of $\mathscr{L}_m$, representing matter sources, introduce additional forces orthogonal to the four-velocity \cite{extraforce1,extraforce2}. Recent studies suggest that $\mathpzc{f}(\mathcal{R},\mathscr{L}_m)$ gravity may provide a viable explanation for several cosmological and astrophysical aspects \cite{frlm1,frlm2,frlm3,frlm4,frlm5,frlm6,frlm7}. In the present paper, our main focus is to study the influence of such coupling on the solution of wormholes with noncommutativity and conformal symmetry. Furthermore, it has been argued that the noncommutative effects can be incorporated by solely manipulating the matter source \cite{intrinsic}. It is quite interesting to study curvature-matter coupling along with noncommutative geometry which would probably lead to the modification of the Einstein tensor part of the field equations as well as the matter sector part.

\subsection{Governing Equations in $\mathpzc{f}(\mathcal{R},\mathscr{L}_m)$ Gravity}

    The modified action is described by \cite{frlm}, 
\begin{equation}\label{action}
			S=\int \mathpzc{f}(\mathcal{R},\mathscr{L}_m)	\sqrt{-g}\, d^4x,
		\end{equation}	
where $\mathpzc{f}$ is an arbitrary function of the scalar curvature $\mathcal{R}$ and the matter Lagrangian $\mathscr{L}_m$. For the specific case $\mathpzc{f}=\mathcal{R}/2+\mathscr{L}_m$, the governing equations of GR are recovered. 
	 It describes the fundamental principles and dynamics of the system. By employing equation \eqref{action}, we can derive the field equations associated with $\mathpzc{f}(\mathcal{R},\mathscr{L}_m)$ gravity. To this end, we vary equation \eqref{action} with respect to $g^{ab}$, resulting in the following expression:
		\begin{equation}\label{fieldequation1}
			\begin{split}
				\mathpzc{f}_\mathcal{R}\mathcal{R}_{ab}+(g_{ab}\nabla_a\nabla^{a}-\nabla_a\nabla_b)\mathpzc{f}_\mathcal{R}-\dfrac{1}{2}\left[\mathpzc{f}- \mathpzc{f}_{\mathscr{L}_m}\mathscr{L}_m\right]g_{ab}\\=\dfrac{1}{2}\mathpzc{f}_{\mathscr{L}_m}\mathcal{T}_{ab},
			\end{split}
		\end{equation}
		where $\mathpzc{f}_{\mathscr{L}_m}$ denotes the derivative with respect to $\mathscr{L}_m$ and $\mathpzc{f}_\mathcal{R}$ represents the derivative with respect to the Ricci scalar $\mathcal{R}$.  Further, the Energy-Momentum tensor (EMT) $\mathcal{T}_{ab}$ is given by,
		\begin{equation}\label{emt}
			\mathcal{T}_{ab}=-\dfrac{2}{\sqrt{-g}} \dfrac{\delta(\sqrt{-g}\mathscr{L}_m)}{\delta g^{ab}}.
		\end{equation}  
		
    The covariant divergence of EMT leads to the expression:
		\begin{equation}\label{divofT}
			\nabla^a \mathcal{T}_{ab}=2\left\lbrace \nabla^a \text{ln}\left[\mathpzc{f}_{\mathscr{L}_m} \right]\right\rbrace \dfrac{\partial \mathscr{L}_m }{\partial g^{ab}}. 
		\end{equation}
		\par Now contracting the governing equation \eqref{fieldequation1} we obtain the following correspondence between matter Lagrangian and the trace of EMT:
		\begin{equation}\label{traceoffieldequation}
			\begin{split}
				3\nabla_a\nabla^{a}\mathpzc{f}_\mathcal{R}+\mathpzc{f}_\mathcal{R}\mathcal{R}-2\left[\mathpzc{f} -\mathpzc{f}_{\mathscr{L}_m}\mathscr{L}_m\right]=\dfrac{1}{2}\mathpzc{f}_{\mathscr{L}_m}\mathcal{T}.
			\end{split}
		\end{equation}
		Using this, the field equation can be rewritten as,
		\begin{equation}\label{fieldquation2}
			\begin{split}
				\mathpzc{f}_\mathcal{R}\left( \mathcal{R}_{ab}-\dfrac{1}{3}\mathcal{R}g_{ab}\right) + \dfrac{g_{ab}}{6}\left[\mathpzc{f} -\mathpzc{f}_{\mathscr{L}_m}\mathscr{L}_m\right]\\=\dfrac{1}{2}\left(\mathcal{T}_{ab} -\dfrac{1}{3}\mathcal{T}g_{ab}\right)\mathpzc{f}_{\mathscr{L}_m}(\mathcal{R},\mathscr{L}_m)+\nabla_a\nabla_{b}\mathpzc{f}_\mathcal{R}.
			\end{split}
		\end{equation}	
        
\subsection{Wormhole Metric and Criteria of Traversability}
   The Morris-Thorne metric for the traversable wormhole is described as \cite{morrisandthorne},
 \begin{equation}\label{whmetric}
		ds^2=e^{2\lambda(r)}dt^2-\dfrac{dr^2}{1-\dfrac{S(r)}{r} }  - r^2\left(d\theta^2+\text{sin}^2\theta \,d\phi^2\right). 
	\end{equation} 

  Here, the functions $\lambda(r)$ and $S(r)$, represent the redshift and shape functions respectively. The redshift function $\lambda$, should be finite throughout the entire spacetime to avoid the presence of a horizon. Along with this, it should vanish for larger domain values in asymptotically flat spacetimes. The radial coordinate $r$ spans from $r_0$ to infinity, with $r_0$ being the throat radius, where the shape function $S(r)$ has a fixed point in order to satisfy the throat condition, specifically $S(r_0) = r_0$. The shape function is significant in ensuring the traversability of the wormhole. It is a monotonically increasing function that asymptotically approaches flat spacetime, as $\frac{S(r)}{r}$ tends to zero for $r\to\infty$. Additionally, the shape function satisfies the flaring-out condition, given by $\frac{S(r) - rS'(r)}{S(r)^2} > 0$, which, at the throat, translates to $S'(r_0) < 1$. Another essential function in describing the geometry of a traversable wormhole is the proper radial distance function which is given by
 \begin{equation}\label{prd}
    L(r)=\pm \int_{r_0}^r \sqrt{\dfrac{r}{r-S(r)}}dr.
 \end{equation}

\subsection{Energy Conditions}

In the realm of wormhole physics, a prominent aspect is the violation of energy conditions. However, it is important to address a nuanced concern that arises in the context of modified theories of gravity, where the gravitational field equations deviate from the conventional relativistic Einstein equations. This concern pertains to the relationship between energy conditions and the field equations. Energy conditions traditionally emerge from the Raychaudhuri equation, which incorporates a term denoted as $\mathcal{R}_{ab}k^ak^b$ (NEC) when considering the expansion of geodesics along any null vector represented by $k^a$. Condition $\mathcal{R}_{ab}k^ak^b\ge0$ is imposed to ensure the convergence of geodesic congruences within a finite range of the parameter. In the framework of general relativity, the energy conditions can be expressed in terms of the stress-energy tensor denoted by $\mathcal{T}_{ab}k^ak^b\ge0$. However, in alternative theories of gravity, the replacement of the Ricci tensor using the corresponding field equations becomes a non-trivial task. While evaluating condition $\mathcal{R}_{ab}k^ak^b$ is straightforward when an Einstein-Hilbert term is present, in the case of modified gravity theories, the process is not as obvious. Thus, to simplify matters, it becomes necessary to express Equation \eqref{fieldquation2} as an effective gravitational field equation. Thus,

\begin{equation}\label{eq:fieldeff}
    G_{ab}= \mathcal{T}_{ab}^{eff},
\end{equation}
where

\begin{equation}
\begin{split}
     \mathcal{T}_{ab}^{eff}=\frac{1}{\mathpzc{f}_{\mathcal{R}}}\left[\dfrac{1}{2}\left(\mathpzc{f}- \mathcal{R}\mathpzc{f}_{\mathcal{R}}\right)g_{ab}-(g_{ab}\nabla_a\nabla^{a}-\nabla_a\nabla_b)\mathpzc{f}_\mathcal{R}\right.\\\left.+\frac{1}{2}\mathpzc{f}_{\mathscr{L}_m}\mathscr{L}_m g_{ab}+\frac{1}{2}\mathpzc{f}_{\mathscr{L}_m}\mathcal{T}_{ab}\right].
\end{split} 
\end{equation}

On contracting the above equation, we get
\begin{equation}
    \mathcal{T}^{eff}=\frac{1}{\mathpzc{f}_{\mathcal{R}}}\left[2\left(\mathpzc{f} -\mathcal{R}\mathpzc{f}_{\mathcal{R}}\right)-3\nabla_a\nabla^{a}\mathpzc{f}_\mathcal{R}+2\mathpzc{f}_{\mathscr{L}_m}\mathscr{L}_m+\dfrac{1}{2}\mathpzc{f}_{\mathscr{L}_m}\mathcal{T}\right].
\end{equation}

Therefore, with the aid of these expressions, one can represent the Ricci tensor as

\begin{equation}
    \mathcal{R}_{ab}=\mathcal{T}_{ab}^{eff}-\frac{1}{2}g_{ab}\mathcal{T}^{eff}.
\end{equation}

Furthermore, for the wormhole metric \eqref{whmetric}, the field equation \eqref{eq:fieldeff} takes the form,

\begin{gather}
    \rho^{eff}=\frac{S'}{r^2},\\
    p_r^{eff}=\frac{S}{r^3}-2\left(1-\frac{S}{r}\right)\frac{\lambda'}{r},\\
    \begin{split}
        p_{\tau}^{eff}=\left(1-\frac{S}{r}\right)\left[\lambda''+(\lambda')^2-\frac{S'r-S}{2r(r-S)}\lambda'\right.\\\left.-\frac{S'-S}{2r^2(r-S)}+\frac{\lambda'}{r}\right].
    \end{split}
\end{gather}

\subsection{Conformal Killing Vector}

By substituting the expression $\mathcal{L}_{\eta}g_{ab}=\Psi(r) g_{ab}$ from equation \eqref{eq:ckv} into equation \eqref{whmetric}, we obtain the following expressions:

\begin{gather}
    \eta^1 \lambda'(r)=\frac{\Psi(r)}{2},\\
    \eta^1 = \frac{r\Psi(r)}{2},\\
    \eta^1 \left(-\frac{S(r)-rS'(r)}{r^2-rS(r)}\right)+2\eta^1_{,1}=\Psi(r).
\end{gather}

Consequently, from the above system of equations, we get,

\begin{equation}\label{eq:ckvsol}
     e^{2\lambda(r)}=K_1 r^2,~\text{and}~
    \left(1-\frac{S(r)}{r}\right)^{-1}=\frac{K_2}{\Psi^2(r)}.
\end{equation}

For the sake of simplicity, we assume $H(r)=\Psi^2(r)$. Subsequently, the expression for the shape function $S(r)$ becomes:

\begin{equation}\label{eq:main}
    S(r)=r\left(1-\frac{H(r)}{K_2}\right).
\end{equation}

\section{Wormhole Models in $\mathpzc{f}(\mathcal{R},\mathscr{L}_m)$ Gravity}\label{III}

		\par We shall now consider a viable form of the curvature-matter coupling, described by    
				\begin{equation}\label{mod}
		    \mathpzc{f}(\mathcal{R},\mathscr{L}_m)=\alpha\dfrac{\mathcal{R}}{2}+(1+\beta\mathcal{R}_0)\mathscr{L}_m,
		\end{equation}
	    where $\alpha$ is a scalar, $\beta$ is the coupling constant and for $\alpha=1,\beta=0$ one can retain GR. $\mathcal{R}_0$ is a constant with the Ricci scalar's value at wormhole throat. The present model is motivated by the generalized formulation of non-minimal coupling, proposed in \cite{frlm} which is given by $\mathpzc{f}(\mathcal{R},\mathscr{L}_m) = f_1(\mathcal{R})+f_2(\mathcal{R})G(\mathscr{L}_m)$, where $f_1$ and $f_2$ are functions of the Ricci scalar $\mathcal{R}$, and $G(\mathscr{L}_m)$ depends on the matter Lagrangian $\mathscr{L}_m$. In our current model, we consider a specific configuration with $f_1(\mathcal{R})=\alpha\frac{\mathcal{R}}{2}$, $f_2(\mathcal{R})=1+\beta \mathcal{R}_0$, and $G(\mathscr{L}_m)=\mathscr{L}_m$. Further, the Lagrangian describes the physical aspects of spacetime and aids in analyzing the motion of test particles. The choice of the matter Lagrangian clearly specifies the matter distribution in spacetime by providing the corresponding energy-momentum candidate. In the literature, numerous works can be found that depend on different choices of $\mathscr{L}_m $ \cite{lmrho}. The energy density-dependent matter Lagrangian, in the case of $\mathscr{L}_m=\rho$ or $-\rho$, results in the energy-momentum tensor with only $\mathcal{T}^t_t$ being non-zero, and the rest of the components vanish. Here, we are interested in studying scenarios in which matter density is anisotropic in nature. For this purpose, selecting Lagrangian density $\mathscr{L}_m$ as the function of average pressure $P$ i.e., $\mathscr{L}_m=P$  is the most suitable choice \cite{lmp1,lmp2,lmp3}.

        For anisotropic matter, the energy-momentum tensor can be represented as,
        \begin{equation}\label{energymomentumtensor}
			\mathcal{T}_{ab}=(\rho+p_\tau)\eta_a \eta_b-p_\tau\,g_{ab}+(p_r-p_\tau)\xi_{a}\xi_b,
        \end{equation}
        where, $\eta^a$ is the 4-velocity, $\xi^{a}=\sqrt{1-S(r)/r} \delta^{a}_r$ is a vector in the direction of radial coordinate and $\rho,p_r,p_\tau$ are respectively the energy density, radial and tangential pressures. In addition, for the wormhole metric \eqref{whmetric}, the Ricci scalar value reads,

        \begin{equation}\label{R}
            \mathcal{R}=2\left(1-\frac{S}{r}\right)\left[\lambda''+(\lambda')^2-\frac{S'}{r(r-S)}-\frac{S' r+3S-4r}{2r(r-S)}\lambda'\right].
        \end{equation}

        Thus, using \eqref{mod}, \eqref{energymomentumtensor} and \eqref{R} the governing field equations can be expressed as follows:
        
        \begin{widetext}
            \begin{gather}
    \begin{split}
         \frac{\alpha  \left(S' \left(1-r \lambda'\right)+(4 r-3 S) \lambda'+2 r (r-S) \left(\lambda''+\lambda'^2\right)\right)}{r}\\=\frac{r (p_r+2 (p_{\tau}+\rho )) \left(\beta  \left(-S'_0 \left(r_0 \lambda'_0+2\right)+(4 r_0-3 S_0) \lambda'_0+2 r_0 (r_0-S_0) \left(\lambda_0''+{\lambda'_0}^{2}\right)\right)+r_0^2\right)}{r_0^2}
    \end{split},
       \\
    \begin{split}
          \alpha  \left(r \left(\lambda ' \left(S'-2 r \lambda '+2\right)-2 r \lambda ''\right)+2 S'+S \left(2 r \lambda ''+\lambda ' \left(2 r \lambda '-3\right)-\frac{3}{r}\right)\right) \\=\frac{r^2 (2 p_r-2 p_{\tau}+\rho ) \left(\beta  \left(-S'_0 \left(r_0 \lambda '_0+2\right)+(4 r_0-3 S_0) \lambda '_0+2 r_0 (r_0-S_0) \left(\lambda ''_0+{\lambda'_0}^2\right)\right)+r_0^2\right)}{r_0^2}
    \end{split},
    \\
    \begin{split}
        \frac{\alpha  \left(r \left(r \lambda'-1\right) \left(S'-2 r \lambda'\right)+S \left(r \left(2 r \lambda ''+\lambda ' \left(2 r \lambda '-3\right)\right)-3\right)-2 r^3 \lambda ''\right)}{2 r}\\=\frac{r^2 (p_r-p_{\tau}-\rho ) \left(\beta  \left(-S'_0 \left(r_0 \lambda '_0+2\right)+(4 r_0-3 S_0) \lambda '_0+2 r_0 (r_0-S_0) \left(\lambda''_0+{\lambda '_0}^2\right)\right)+r_0^2\right)}{r_0^2}
    \end{split}.
    \end{gather}
        \end{widetext}

Here, primes ($'$) represent the derivative with respect to $r$, $r_0$ is the throat radius and $S'_0$, $\lambda'_0$ are the derivatives of the shape function and redshift function at the wormhole throat.  By employing relations \eqref{eq:ckvsol}, \eqref{eq:main} and adopting dimensionless parameters, the set of equations can be solved to obtain the values of $\rho$, $p_r$, and $p_\tau$. These are given by,

\begin{widetext}
    \begin{gather}
    \label{eq:ode}\rho_{*} \left(\frac{r}{\sqrt{\Theta }}\right)=-\frac{\alpha   \frac{r_0^2}{\Theta} \left[\frac{r}{\sqrt{\Theta }}\dot{H}_{*}\left(\frac{r}{\sqrt{\Theta }}\right)+H_{*}\left(\frac{r}{\sqrt{\Theta }}\right)-K_2\right]}{\frac{r^2}{\Theta}  \left[3 \Tilde{\beta}  \frac{r_0}{\sqrt{\Theta }} \dot{H}_{*}\left(\frac{r_0}{\sqrt{\Theta }}\right)+6 \Tilde{\beta}  H_{*}\left(\frac{r_0}{\sqrt{\Theta }}\right)+K_2 \left(\frac{r_0^2}{\Theta} -2 \Tilde{\beta} \right)\right]},\\
    \label{eq:pr}p_{r*}\left(\frac{r}{\sqrt{\Theta }}\right)=-\frac{\alpha  \frac{r_0^2}{\Theta} \left[K_2-3 H_{*}\left(\frac{r}{\sqrt{\Theta }}\right)\right]}{\frac{r^2}{\Theta}  \left[3 \Tilde{\beta}  \frac{r_0}{\sqrt{\Theta }} \dot{H}_{*}\left(\frac{r_0}{\sqrt{\Theta }}\right)+6 \Tilde{\beta}  H_{*}\left(\frac{r_0}{\sqrt{\Theta }}\right)+K_2 \left(\frac{r_0^2}{\Theta} -2 \Tilde{\beta} \right)\right]},\\
     \label{eq:pt}p_{\tau *}\left(\frac{r}{\sqrt{\Theta }}\right)=\frac{\alpha   \frac{r_0^2}{\Theta} \left[\frac{r}{\sqrt{\Theta }}\dot{H}_{*}\left(\frac{r}{\sqrt{\Theta }}\right)+H_{*}\left(\frac{r}{\sqrt{\Theta }}\right)\right]}{\frac{r^2}{\Theta}  \left[3 \Tilde{\beta}  \frac{r_0}{\sqrt{\Theta }} \dot{H}_{*}\left(\frac{r_0}{\sqrt{\Theta }}\right)+6 \Tilde{\beta}  H_{*}\left(\frac{r_0}{\sqrt{\Theta }}\right)+K_2 \left(\frac{r_0^2}{\Theta} -2 \Tilde{\beta} \right)\right]},
\end{gather}
\end{widetext}

where subscript `$_*$' represents corresponding adimensional quantities and the overhead dot is the derivative of the function with respect to $\frac{r}{\sqrt{\Theta }}$. The inclusion of noncommutative geometry becomes more prominent in regions close to the origin. The minimal length $\sqrt{\Theta}$ plays a crucial role in describing the plausible neighborhood where noncommutativity regularizes both radial and tangential pressures, as well as matter density. This makes the region $r \lesssim \Theta$ of particular interest in our study. However, one cannot fix $\Theta$ by a particular choice. To address this issue, we reparameterize the quantities during the calibration process, moving from dimensional to non-dimensional representations. This approach offers the added advantage of avoiding complexities that can arise when dealing with parameters of different dimensions in field equations.

\subsection{Gaussian energy density}

In this subsection, we shall focus on the Gaussian noncommutative geometry (GNC) of the particle-like gravitational sources. By plugging in the GNC energy density \eqref{gauss} in equation \eqref{eq:ode}. we get the differential equation,

\begin{widetext}
     \begin{equation}\label{eq:GNCode}
       -\frac{\alpha   \frac{r_0^2}{\Theta} \left[\frac{r}{\sqrt{\Theta }}\dot{H}_{*}\left(\frac{r}{\sqrt{\Theta }}\right)+H_{*}\left(\frac{r}{\sqrt{\Theta }}\right)-K_2\right]}{\frac{r^2}{\Theta}  \left[3 \Tilde{\beta}  \frac{r_0}{\sqrt{\Theta }} \dot{H}_{*}\left(\frac{r_0}{\sqrt{\Theta }}\right)+6 \Tilde{\beta}  H_{*}\left(\frac{r_0}{\sqrt{\Theta }}\right)+K_2 \left(\frac{r_0^2}{\Theta} -2 \Tilde{\beta} \right)\right]}=\frac{\Tilde{M}e^{ -\frac{r^2}{4\Theta}}}{8 \pi ^{3/2}}.
   \end{equation}
\end{widetext}

   Here, $\Tilde{M}$ and $\Tilde{\beta}$ have the dimension of $\frac{r}{\sqrt{\Theta}}$ and are given by $\frac{M}{\sqrt{\Theta}}$ and $\frac{\beta}{\Theta}$, respectively. Clearly, the differential equation above constitutes an initial value problem. In order to analyze the behavior of the shape function, it is imperative to derive the solution of \eqref{eq:GNCode} in terms of the function $H_*$ and subsequently utilize the relation \eqref{eq:main}. To establish the initial value for the function $H_*$, we apply the condition based on $S_*$. It is interesting to note that for the shape function $S_*$ to satisfy the throat condition, we impose the initial condition $H_*\left(\frac{r_0}{\sqrt{\Theta }}\right)=0$ using the relation \eqref{eq:main}. Further, taking $\dot{H}_{*}\left(\frac{r_0}{\sqrt{\Theta }}\right)=n$, for $n$ being real, the particular solution of the initial value problem \eqref{eq:GNCode} can be expressed as, 

    \begin{widetext}
        \begin{equation}
   \begin{split}
        H_*\left(\frac{r}{\sqrt{\Theta }}\right)=&\frac{1}{4 \pi ^{3/2} \alpha  \frac{r}{\sqrt{\Theta }} \frac{r_0^2}{\Theta}} \left[\sqrt{\pi } \Tilde{M} \left(\text{erf}\left(\frac{r_0}{2\sqrt{\Theta }}\right)-\text{erf}\left(\frac{r}{2\sqrt{\Theta }}\right)\right) \left(K_2 \left(\frac{r_0^2}{\Theta}-2 \Tilde{\beta}\right)+3 \Tilde{\beta} n \frac{r_0}{\sqrt{\Theta}}\right)\right.\\&\left.+e^{\frac{1}{4} \left(-\frac{r^2}{\Theta}-\frac{r_0^2}{\Theta}\right)} \left\{e^{\frac{r^2}{4\Theta}} \frac{r_0}{\sqrt{\Theta}} \left(4 \pi ^{3/2} \alpha  K_2 e^{\frac{r_0^2}{4\Theta}} \frac{r_0}{\sqrt{\Theta}} \left(\frac{r}{\sqrt{\Theta}}-\frac{r_0}{\sqrt{\Theta}}\right)-\Tilde{M} \left(K_2 \left(\frac{r_0^2}{\Theta}-2 \Tilde{\beta}\right)+3 \Tilde{\beta} n \frac{r_0}{\sqrt{\Theta}}\right)\right)\right.\right.\\&\left.\left.+\Tilde{M} \frac{r}{\sqrt{\Theta}} e^{\frac{\frac{r_0^2}{\Theta}}{4}} \left(K_2 \left(\frac{r_0^2}{\Theta}-2 \Tilde{\beta}\right)+3 \Tilde{\beta} n \frac{r_0}{\sqrt{\Theta}}\right)\right\}\right].
   \end{split}
   \end{equation}
    \end{widetext}

    With the aid of the above equation, solving $\dot{H}_{*}\left(\frac{r}{\sqrt{\Theta }}\right)$ for $n$, one can find the value of $n$ as,

    \begin{equation}\label{eq:Gn}
        n=\frac{K_2 \sqrt{\Theta}}{r_0}\left(1-\frac{\Tilde{M} \left(\Tilde{\beta}+\frac{r_0^2}{\Theta}\right)}{3 \Tilde{\beta} \Tilde{M}+8 \pi ^{3/2} \alpha  e^{\frac{r_0^2}{4\Theta}}}\right).
    \end{equation}

     \begin{figure*}[t]
            \centering
            \subfloat[$S_*\left(\frac{r}{\sqrt{\Theta}}\right)>0$\label{fig:Gsf1}]{\includegraphics[width=0.3\linewidth]{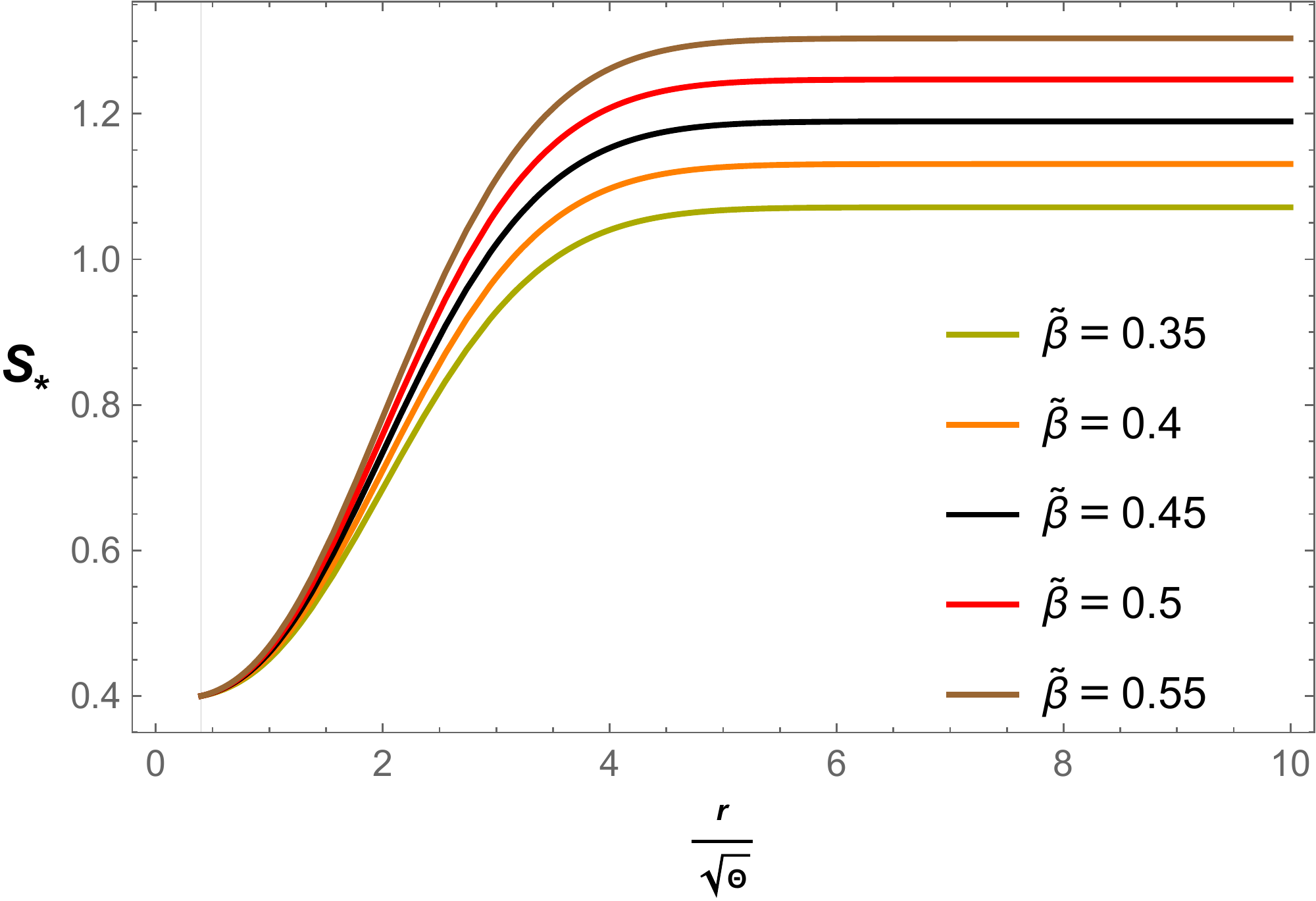}}
    	    \subfloat[$S_*-\frac{r}{\sqrt{\Theta}}<0$\label{fig:Gsf2a}]{\includegraphics[width=0.31\linewidth]{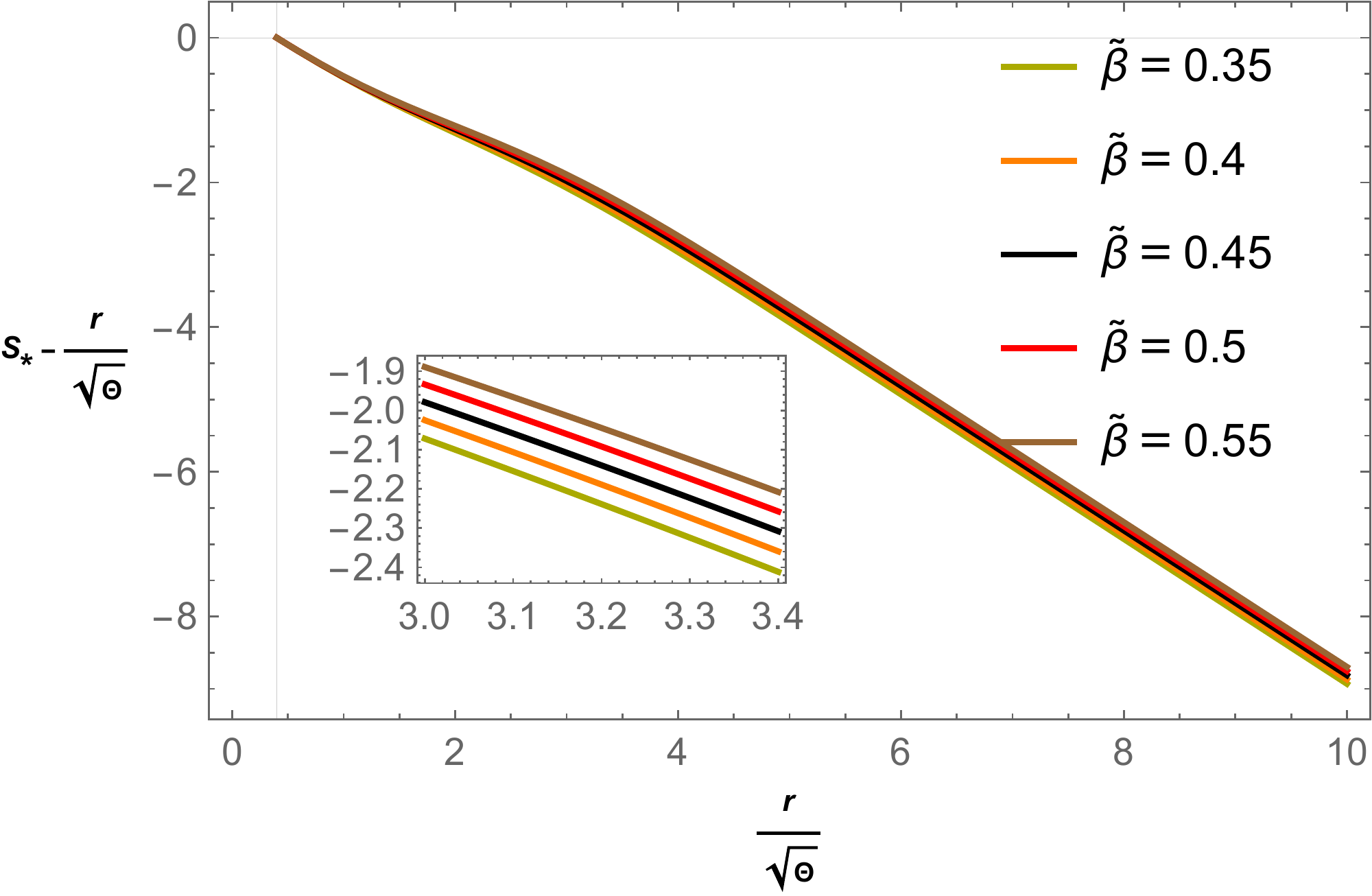}}
    	    \subfloat[$\dot{S}_*\left(\frac{r}{\sqrt{\Theta}}\right)<1$\label{fig:Gsf2b}]{\includegraphics[width=0.3\linewidth]{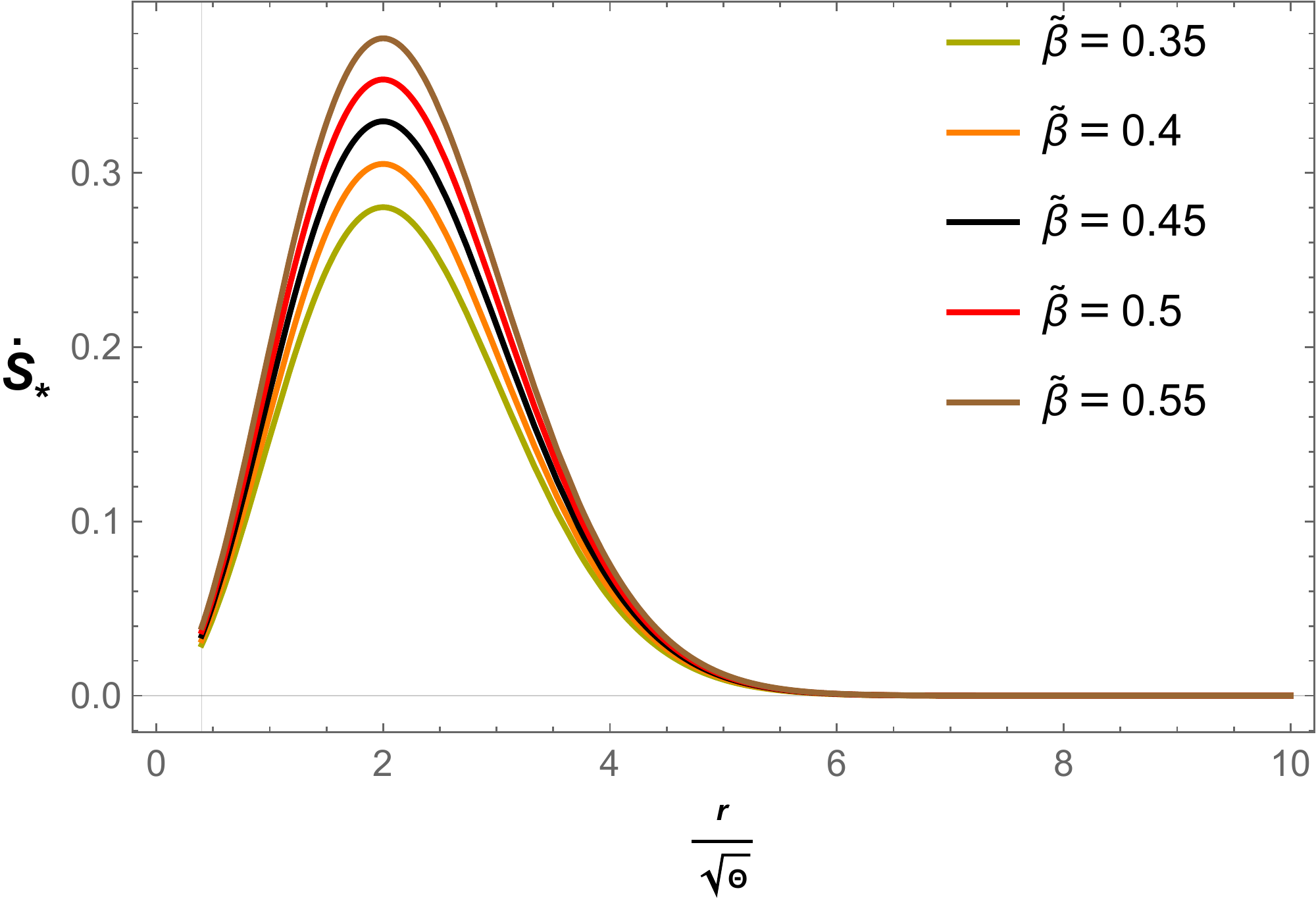}}\\
            \subfloat[$\frac{S_*\left(\frac{r}{\sqrt{\Theta}}\right)-\frac{r}{\sqrt{\Theta}} \dot{S}_*\left(\frac{r}{\sqrt{\Theta}}\right)}{S_*\left(\frac{r}{\sqrt{\Theta}}\right)^2}>0$\label{fig:Gsf3}]{\includegraphics[width=0.3\linewidth]{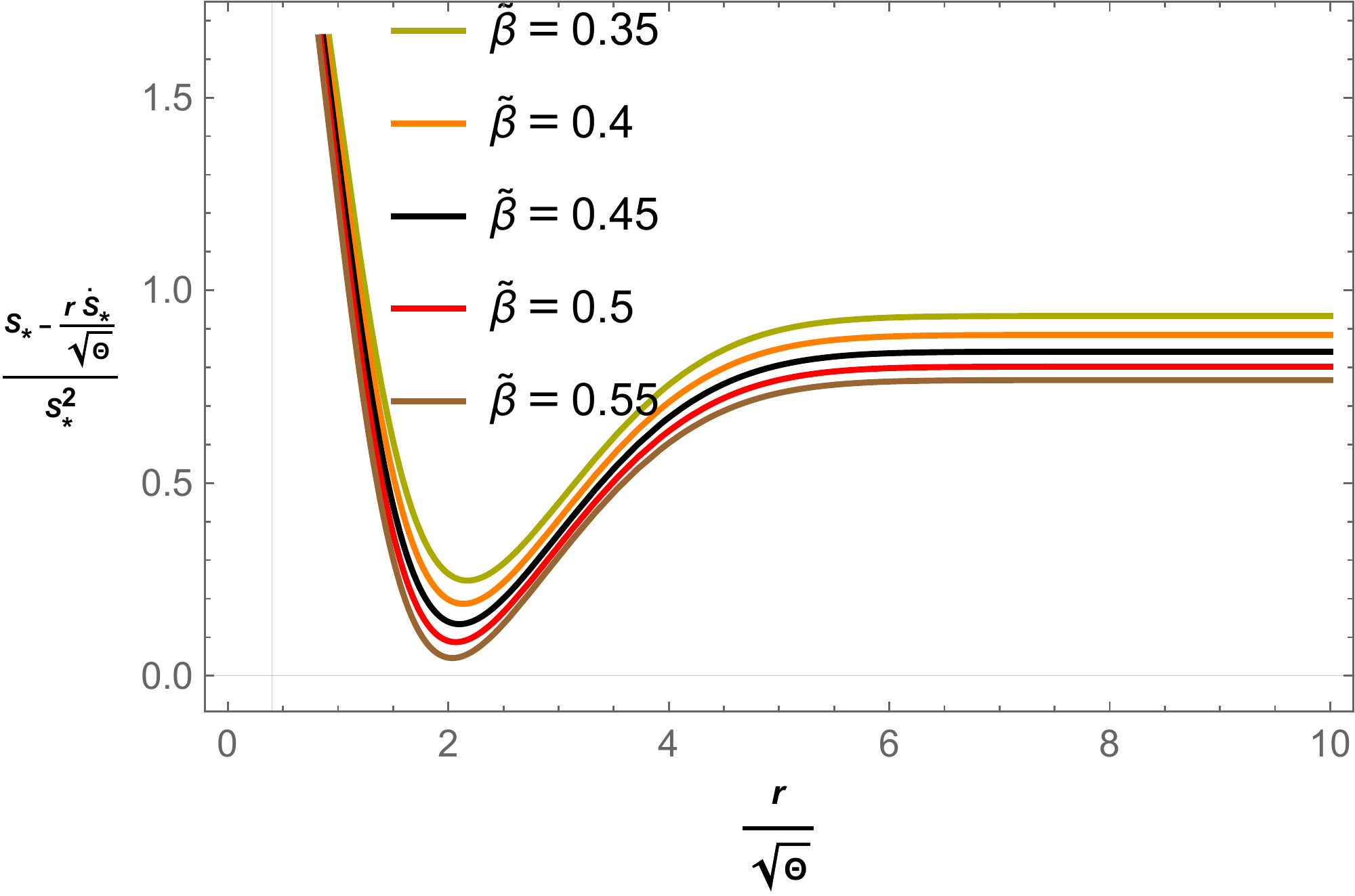}}
            \subfloat[$\frac{S_*\left(\frac{r}{\sqrt{\Theta}}\right)}{\frac{r}{\sqrt{\Theta}}}\to0 $ as $\frac{r}{\sqrt{\Theta}}\to \infty$\label{fig:Gsf4}]{\includegraphics[width=0.3\linewidth]{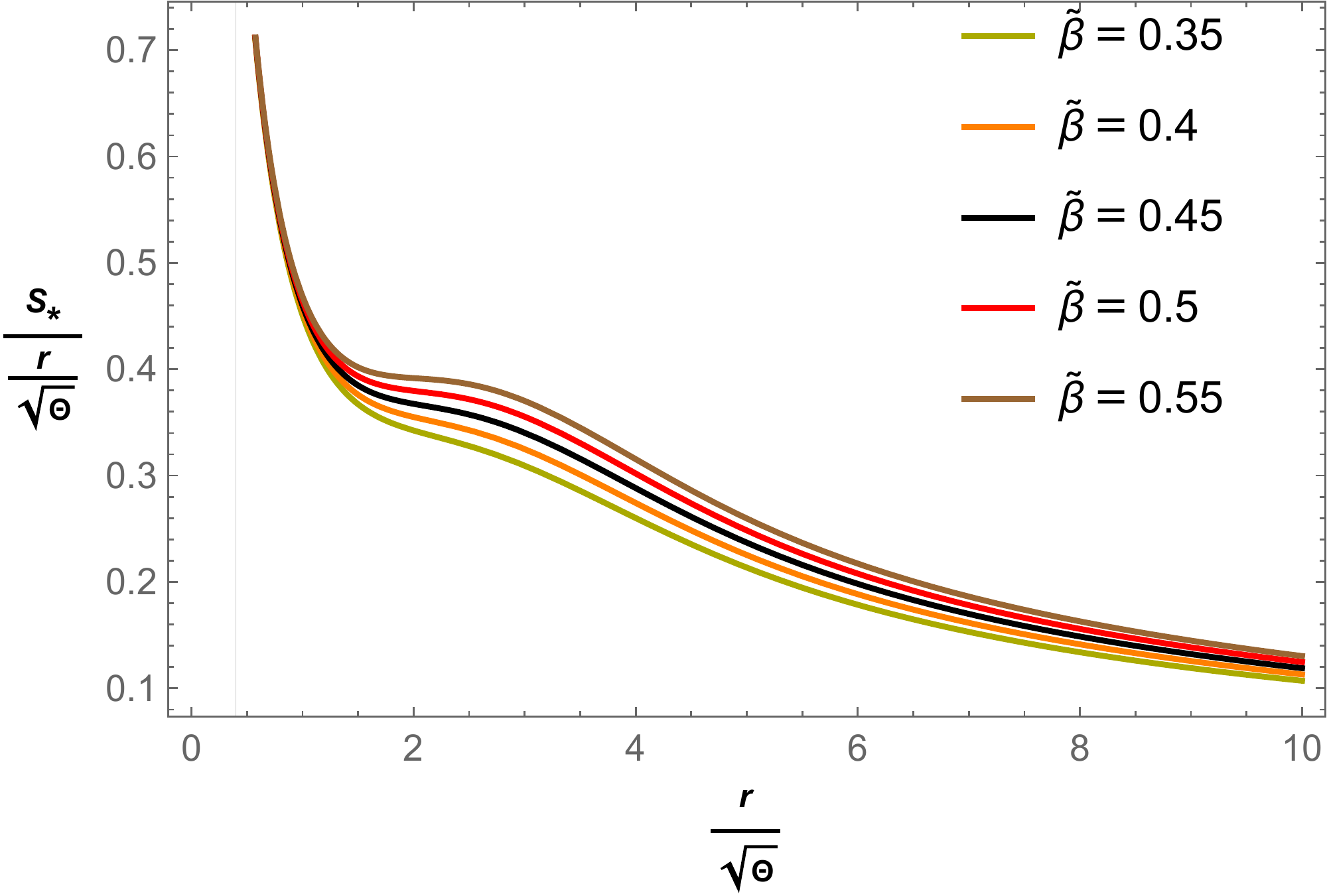}}
            \caption{GNC: Behavior of $S_*$ versus $r/\sqrt{\Theta}$ for different values of $\Tilde{\beta}$ with $\Tilde{M}=3.4$, $\alpha=1.2$, $K_2=2$ and $\frac{r_0}{\sqrt{\Theta}}=0.4$. }
            \label{fig:Gsf}
        \end{figure*}

\begin{figure*}[!]
	    \centering
	    \subfloat[Energy density $\rho_*$\label{fig:Grho}]{\includegraphics[width=0.3\linewidth]{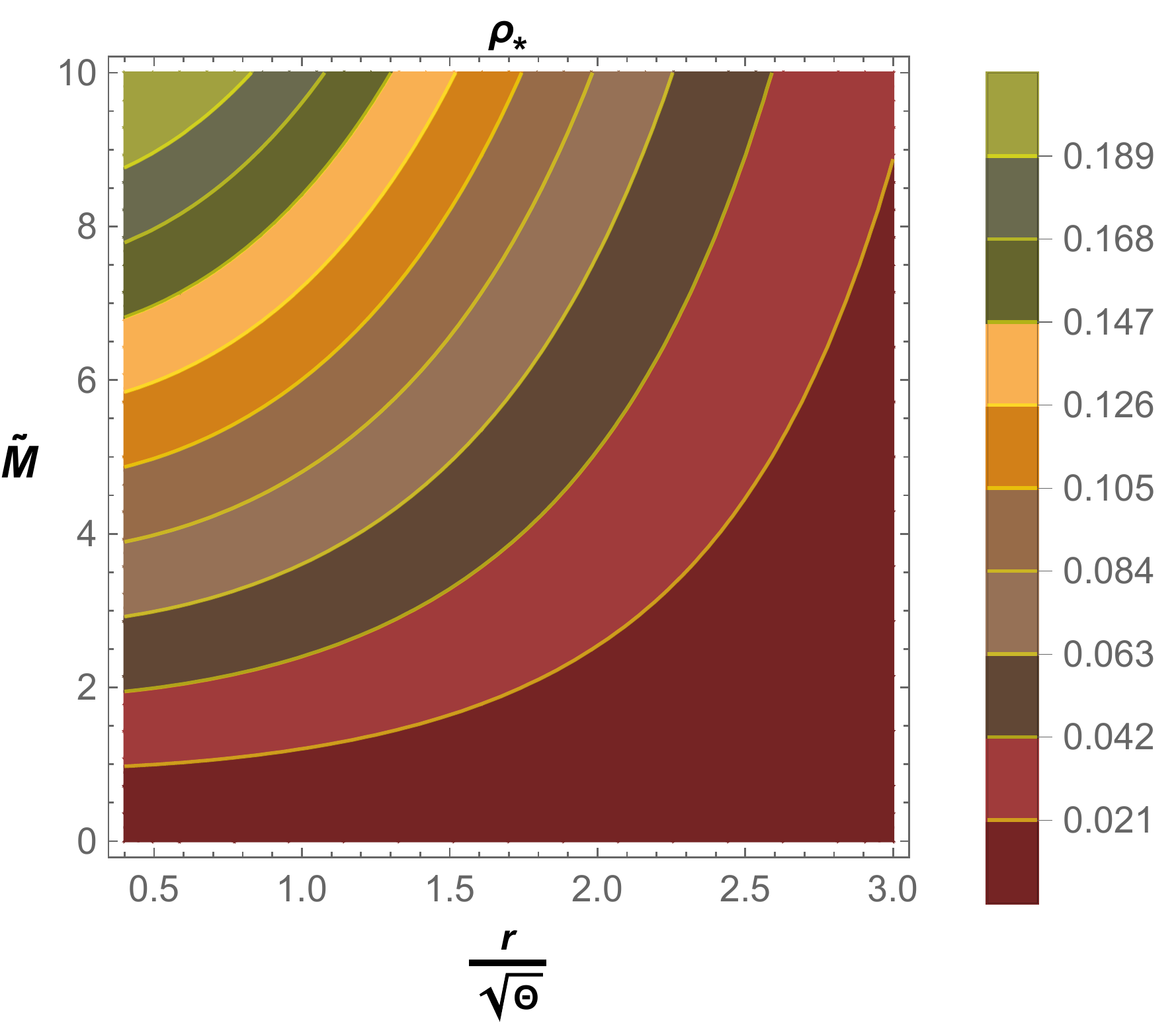}}
	    \subfloat[NEC $\rho_*+p_{r*}$\label{fig:Ge1}]{\includegraphics[width=0.3\linewidth]{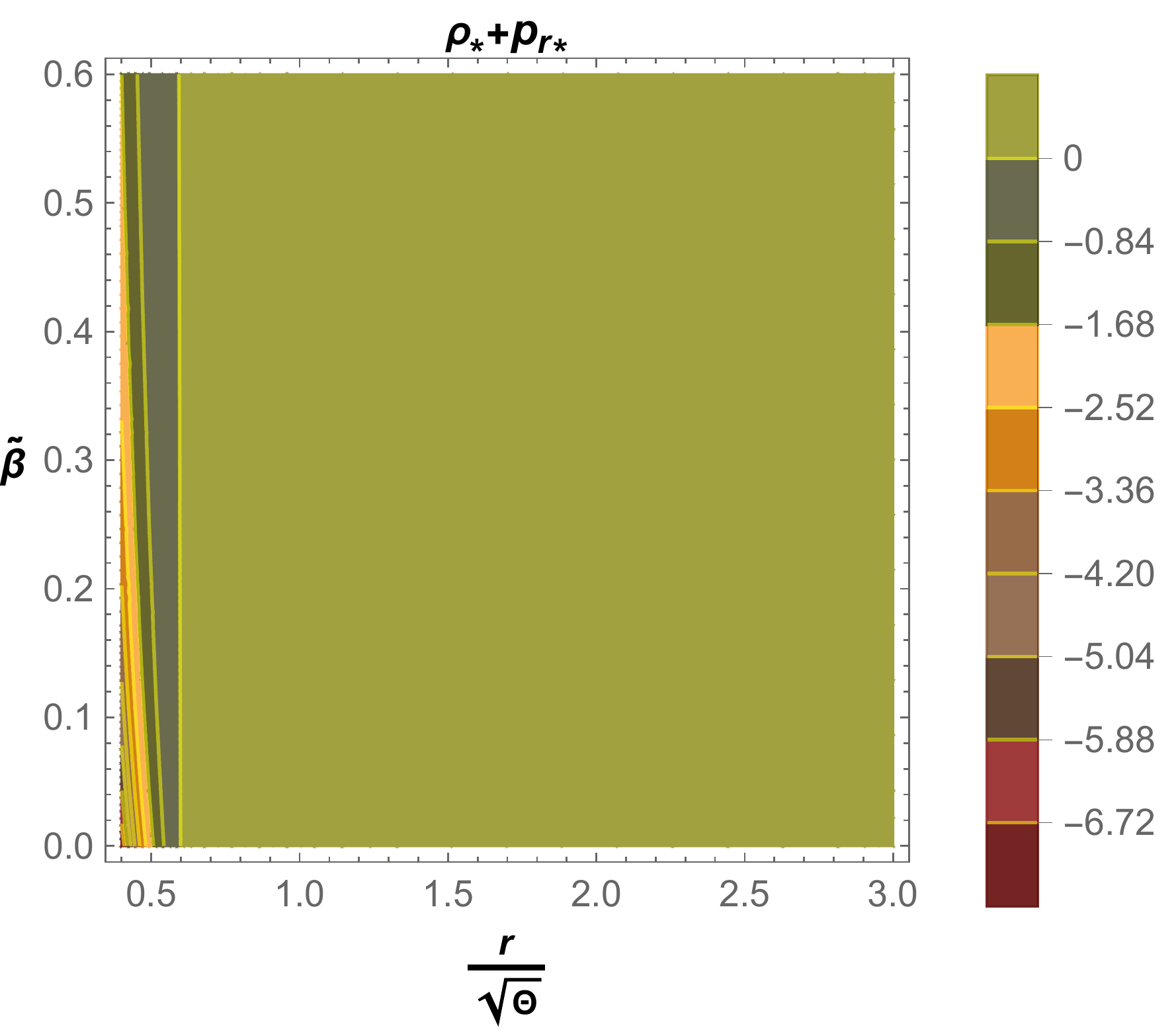}}
	    \subfloat[NEC $\rho_*+p_{\tau*}$\label{fig:Ge2}]{\includegraphics[width=0.3\linewidth]{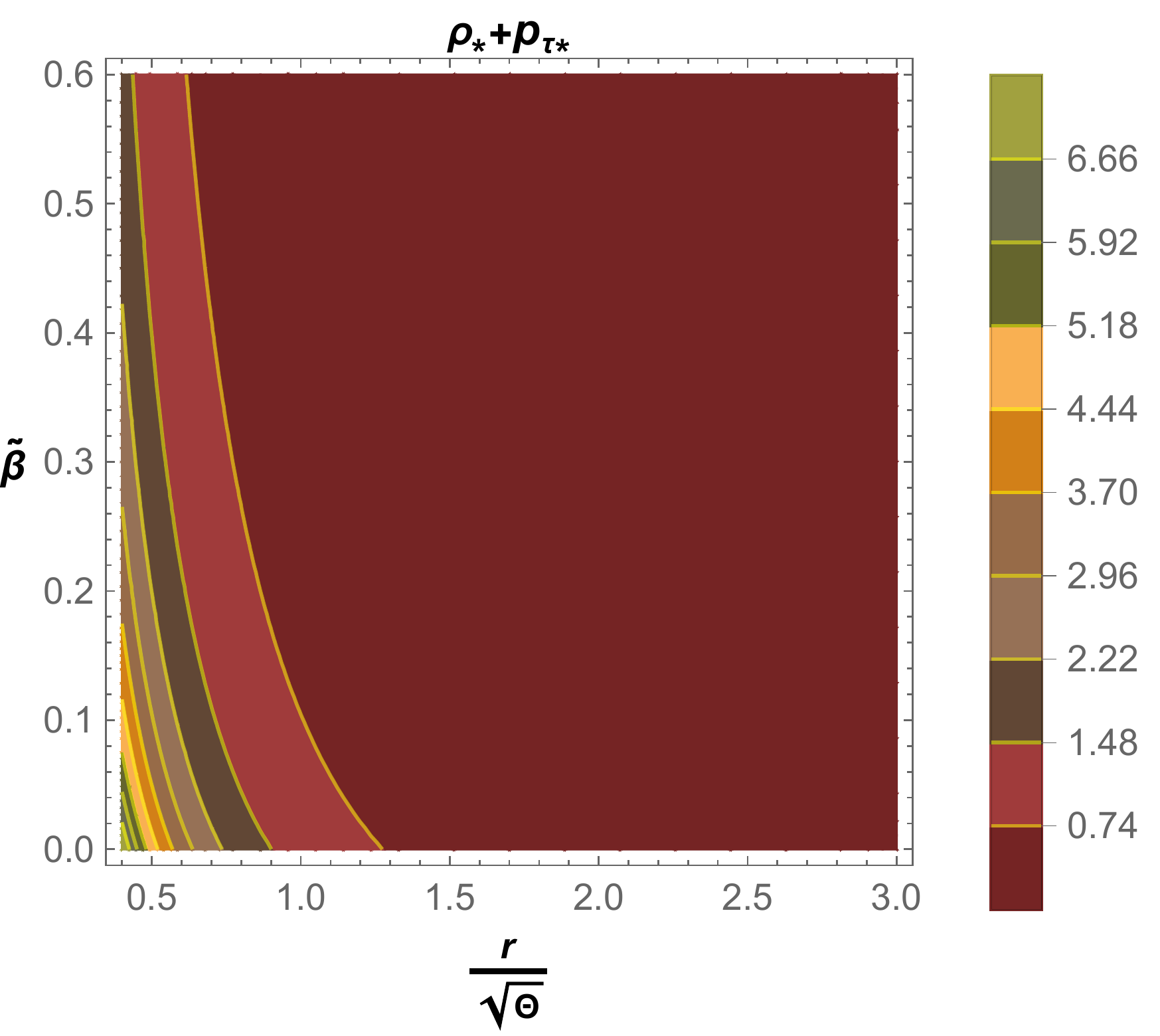}}\\
	    \subfloat[DEC $\rho_*-|p_{r*}|$\label{fig:Ge3}]{\includegraphics[width=0.3\linewidth]{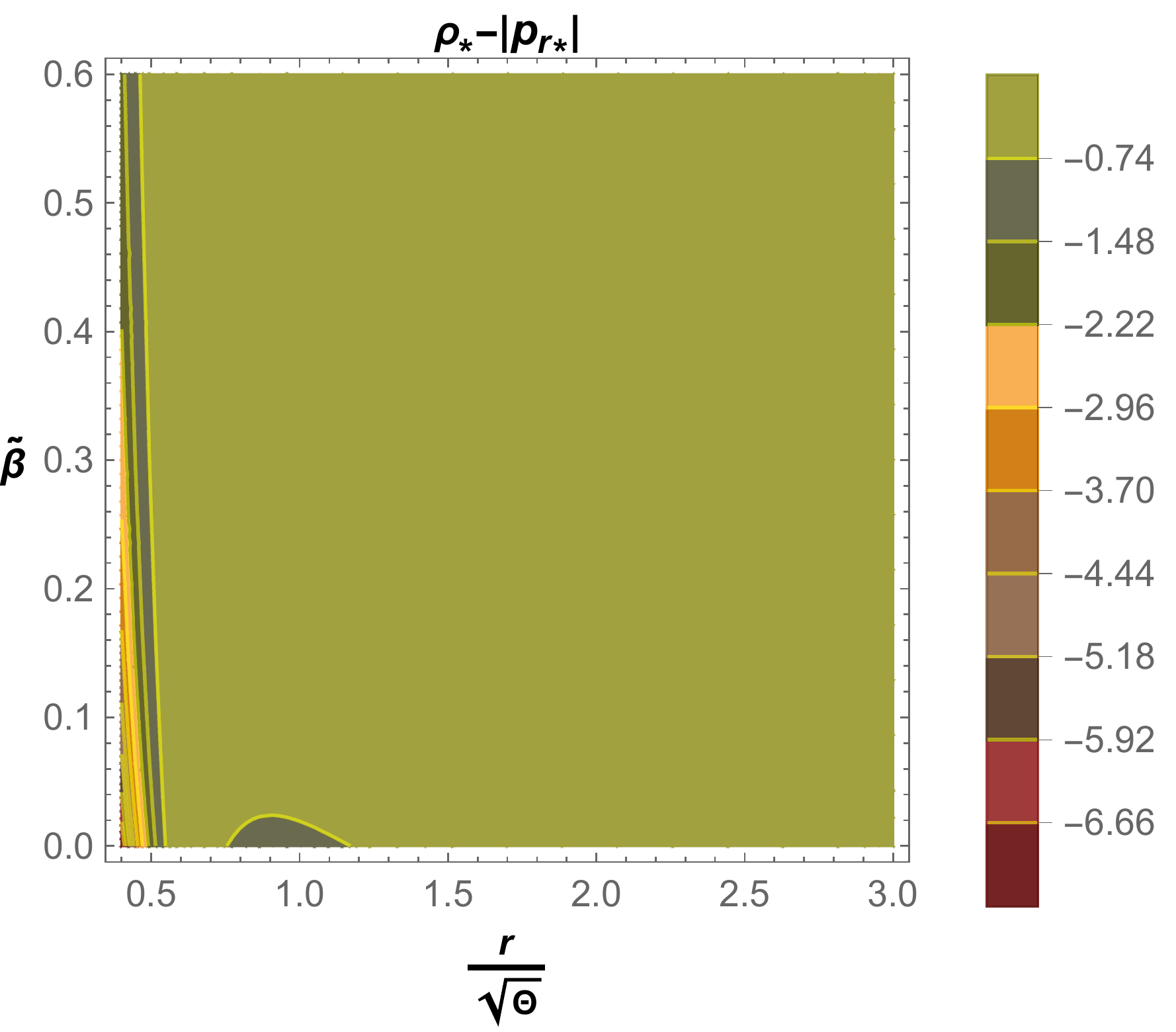}}
	    \subfloat[DEC $\rho_*-|p_{\tau*}|$\label{fig:Ge4}]{\includegraphics[width=0.3\linewidth]{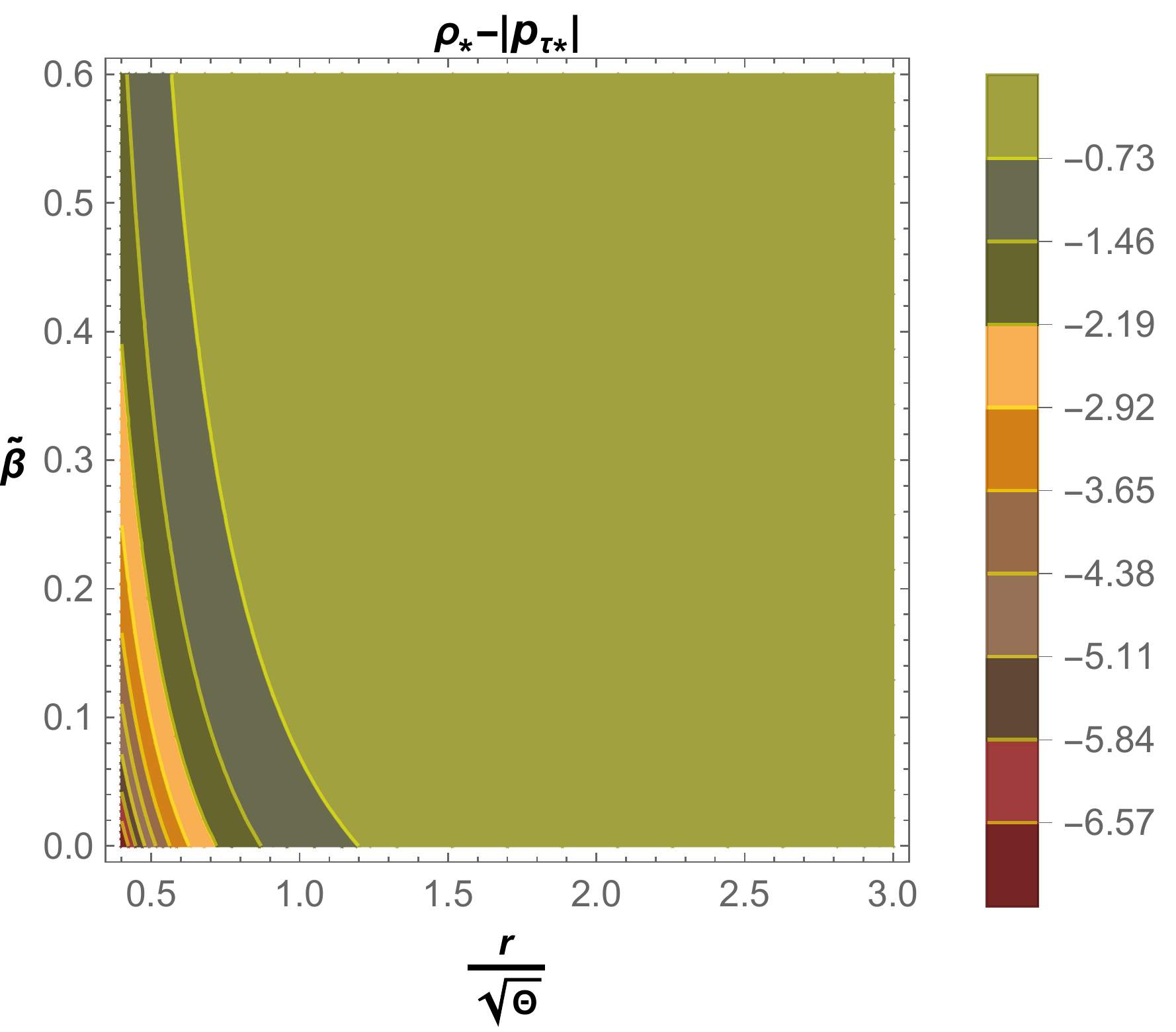}}
	    \subfloat[SEC $\rho_*+p_{r*}+2p_{\tau*}$\label{fig:Ge5}]{\includegraphics[width=0.3\linewidth]{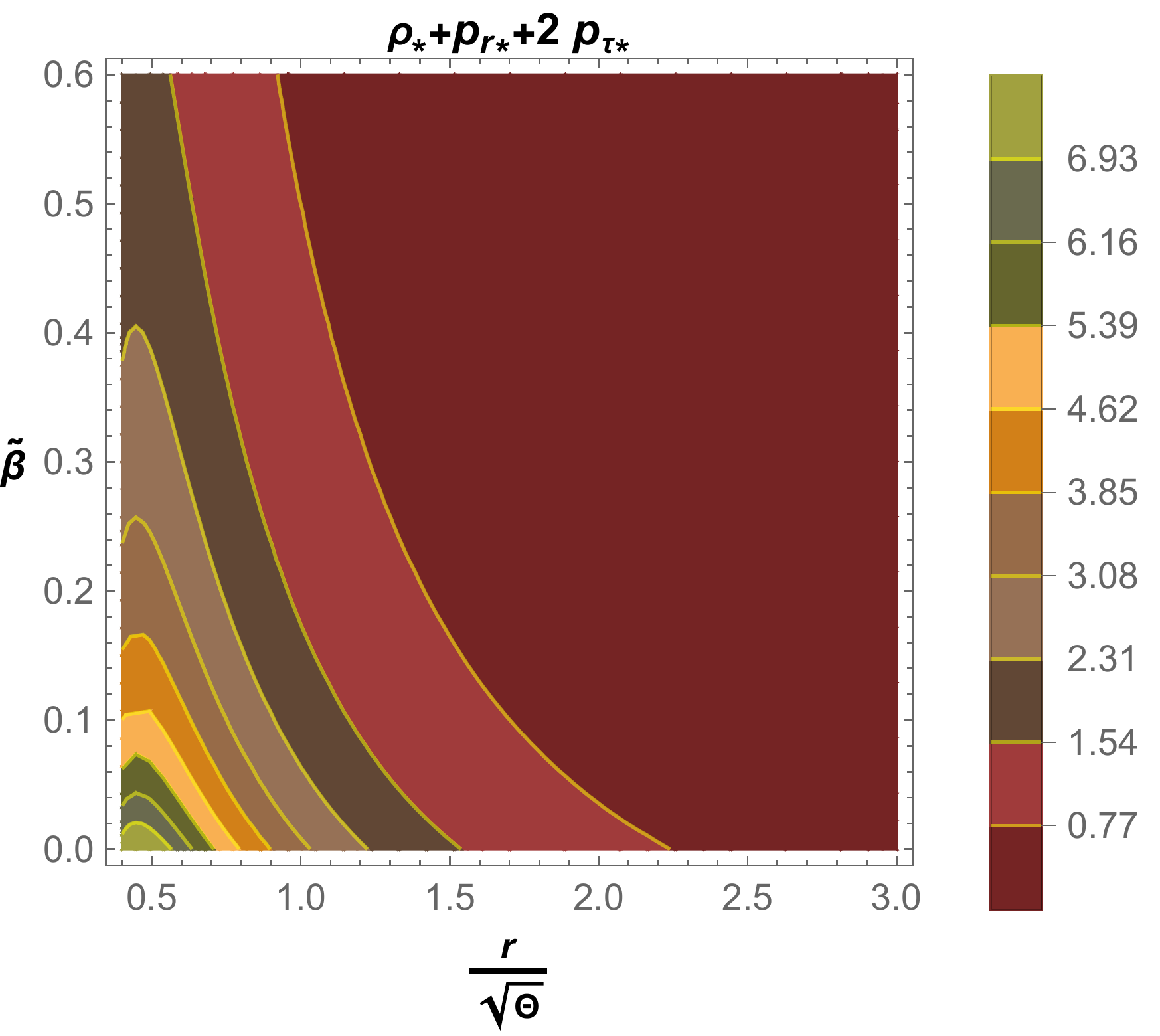}}
	    \caption{GNC: profile of (a) energy density $\rho_*$ varying w.r.t $r/\sqrt{\Theta}$ and $\Tilde{M}$, (b)-(f) different energy conditions varying w.r.t $r/\sqrt{\Theta}$ and model parameter $\Tilde{\beta}$ with $\Tilde{M}=3.4$, $\alpha=1.2$, $K_2=2$ and $\frac{r_0}{\sqrt{\Theta}}=0.4$. }
	    \label{fig:Gec}
	\end{figure*}

    Thus, the corresponding shape function obtained is,

   \begin{widetext}
       \begin{equation}\label{eq:shapefunction}
    \begin{split}
         S_*\left(\frac{r}{\sqrt{\Theta}}\right)=\frac{e^{-\frac{r^2}{4\Theta}} }{\frac{r_0^2}{\Theta} \left(3 \Tilde{\beta} \Tilde{M}+8 \pi ^{3/2} \alpha  e^{\frac{r_0^2}{4\Theta}}\right)}&\left[ \left\{2 \sqrt{\pi } e^{\frac{r_0^2}{4\Theta}} \left(\Tilde{M} \left(\Tilde{\beta}+\frac{r_0^2}{\Theta}\right) \left(\text{erf}\left(\frac{r}{2\Theta}\right)-\text{erf}\left(\frac{r_0}{2\Theta}\right)\right)+4 \pi  \alpha \frac{r_0^3}{\Theta^{\frac{3}{2}}}\right)\right.\right.\\&\left.\left.+\Tilde{M} \frac{r_0}{\sqrt{\Theta}} \left(2 \Tilde{\beta}+(3 \Tilde{\beta}+2) \frac{r_0^2}{\Theta}\right)\right\}e^{\frac{r^2}{4\Theta}}-2 \Tilde{M} \frac{r}{\sqrt{\Theta}} e^{\frac{r_0^2}{4\Theta}} \left(\Tilde{\beta}+\frac{r_0^2}{\Theta}\right)\right].
    \end{split}
    \end{equation}
   \end{widetext}

    Here, the fulfillment of the throat condition can be easily confirmed by the straightforward calculation of $S_*\left(\frac{r_0}{\sqrt{\Theta }}\right)$. Additionally, through an evaluation of the derivative of the shape function $S_*$ at the throat, as indicated by equation \eqref{eq:shapefunction}, we obtain the subsequent relation:
    \begin{equation}
        \dot{S}_{*}\left(\frac{r_0}{\sqrt{\Theta }}\right)=1-\frac{r_0}{\sqrt{\Theta}}\left(\frac{n}{K_2}\right).
    \end{equation}
    
    Thus, to satisfy the flaring-out condition at the wormhole throat, the inequality $1-\frac{r_0}{\sqrt{\Theta}}\left(\frac{n}{K_2}\right)<1$ should be obeyed. Since $r_0\ne 0$ and assuming $K_2>0$ we have $n>0$. Consequently, by referring to equation \eqref{eq:Gn}, we obtain the constraining relation

    \begin{equation}\label{eq:inequalityGNC}
        \frac{\Tilde{M} \left(\Tilde{\beta}+\frac{r_0^2}{\Theta }\right)}{3 \Tilde{\beta} \Tilde{M}+8 \pi ^{3/2} \alpha  e^{\frac{r_0^2}{4\Theta}}}<1.
    \end{equation}

    For GR, the coupling constant vanishes so that the above inequality takes the form,  $\frac{\Tilde{M} \left(\frac{r_0^2}{\Theta } \right)}{8 \pi ^{3/2} e^{\frac{r_0^2}{4\Theta}}}<1$. This expression is significant as it relates the parameters $M, \Theta$, and $r_0$. Moreover, the inequality \eqref{eq:inequalityGNC} conveys that the effective NEC is violated at the throat. Based on these criteria, we carefully select parameter values that satisfy these conditions. For this purpose, we consider $\Tilde{M}=3.4$, $\alpha=1.2$, and $K_2=2$. Furthermore, we take $\frac{r_0}{\sqrt{\Theta}}=0.4$. The primary advantage of this choice is, the influence of noncommutativity becomes prominent in a localized region close to the origin, particularly when the radial coordinate $r$ is smaller than or approximately equal to $\Theta$. In this vicinity, noncommutativity plays a crucial role in regulating the radial and tangential pressures, as well as the density of matter. To this end, $r_0/\sqrt{\Theta}<1$ helps us to achieve this scenario. Additionally, in the present analysis, we intend to examine the influence of the coupling constant $\Tilde{\beta}$ on the behavior of the shape function.

    In \figureautorefname~\ref{fig:Gsf}, we have plotted the profile of the derived shape function $S_*\left(\frac{r}{\sqrt{\Theta}}\right)$ and its characteristic features. Clearly, $S_*\left(\frac{r}{\sqrt{\Theta}}\right)$ is a monotonically increasing function with $S_*\left(\frac{r}{\sqrt{\Theta}}\right)>0$ $\forall~\frac{r}{\sqrt{\Theta}}>\frac{r_0}{\sqrt{\Theta}}$ [\figureautorefname~\ref{fig:Gsf1}], implying $S(r)>0$ $\forall~r>r_0$. Along with this, the wormhole possesses a finite proper radial distance function within the parameter space. This is characterized by the condition  $S_*\left(\frac{r}{\sqrt{\Theta}}\right)<\frac{r}{\sqrt{\Theta}}$ for all $\frac{r}{\sqrt{\Theta}}$ and $\Tilde{\beta} \in [0,3.78)$ (see Figure \ref{fig:Gsf2a}). In addition, from \figureautorefname~\ref{fig:Gsf2b} we have the derivative $\dot{S}_*\left(\frac{r}{\sqrt{\Theta}}\right)$ less than 1, indicating the satisfying behavior of the flaring-out condition at the wormhole throat. Furthermore, for $0\le\Tilde{\beta}<0.6$, the condition $\frac{S_*\left(\frac{r}{\sqrt{\Theta}}\right)-\frac{r}{\sqrt{\Theta}} \dot{S}_*\left(\frac{r}{\sqrt{\Theta}}\right)}{S_*\left(\frac{r}{\sqrt{\Theta}}\right)^2}>0$ holds in the entire domain (\figureautorefname~\ref{fig:Gsf3}), as a consequence of which violation of effective NEC (i.e., $\rho_{eff}+p_{r_{eff}}<0$) is confirmed. We know that the asymptotic condition holds if $\frac{S}{r}\to 0$ as $r\to\infty$. Since minimal length $\sqrt{\Theta}<\infty$, the condition is equivalent to $\frac{S_*}{r/\sqrt{\Theta}}\to0$ for extremely large $r/\sqrt{\Theta}$. In \figureautorefname~\ref{fig:Gsf4}, the plot represents the satisfaction of asymptotic condition for our model.
    On the basis of these observations, we have effectively constrained the parameter space of the dimensionless coupling constant $\Tilde{\beta}$ to the interval $[0,0.6)$, so that $S_*$ satisfies all the required conditions.

    Now, by inserting the obtained $H_*$ function in \eqref{eq:pr} and \eqref{eq:pt}, we get the pressure elements as, 

    \begin{widetext}
         \begin{gather}
  \begin{split}
       p_{r*}=\frac{e^{\frac{1}{4\Theta} \left(-r^2-r_0^2\right)} }{8 \pi ^{\frac{3}{2}} \frac{r^3}{\Theta^{\frac{3}{2}}} \left(\Tilde{\beta}+\frac{r_0^2}{\Theta}\right)}&\left[ \left\{2 \sqrt{\pi } e^{\frac{r_0^2}{4\Theta}} \left(3 \Tilde{M} \left(\Tilde{\beta}+\frac{r_0^2}{\Theta}\right) \left(\text{erf}\left(\frac{r_0}{2\Theta}\right)-\text{erf}\left(\frac{r}{2\Theta}\right)\right)+4 \pi  \alpha  \frac{r_0^2}{\Theta} \left(2 \frac{r}{\sqrt{\Theta}}-3 \frac{r_0}{\sqrt{\Theta}}\right)\right)\right.\right.\\&\left.\left.-3 \Tilde{M} \frac{r_0}{\sqrt{\Theta}} \left(2 \Tilde{\beta}-2 \Tilde{\beta} \frac{r r_0}{\Theta}+(3 \Tilde{\beta}+2) \frac{r_0^2}{\Theta}\right)\right\}e^{\frac{r^2}{4\Theta}}+6 \Tilde{M} \frac{r}{\sqrt{\Theta}} e^{\frac{r_0^2}{4\Theta}} \left(\Tilde{\beta}+\frac{r_0^2}{\Theta}\right)\right],
  \end{split}
     \\
      p_{\tau*}=-\frac{\Tilde{M} e^{-\frac{r^2}{4\Theta}} \frac{r^2}{\Theta} \left(\Tilde{\beta}+\frac{r_0^2}{\Theta}\right)-3 \Tilde{\beta} \Tilde{M} e^{-\frac{r_0^2}{4\Theta}} \frac{r_0^2}{\Theta}-8 \pi ^{3/2} \alpha  \frac{r_0^2}{\Theta}}{8 \pi ^{3/2} \frac{r^2}{\Theta} \left(\Tilde{\beta}+\frac{r_0^2}{\Theta}\right)}.
  \end{gather}
    \end{widetext}

   In our study, it is found that both DECs and radial NEC are violated, whereas tangential NEC and SEC are obeyed. The behavior of ECs along with the energy density profile is illustrated in \figureautorefname~\ref{fig:Gec}.

   \begin{figure*}[!]
            \centering
            \subfloat[$S_*\left(\frac{r}{\sqrt{\Theta}}\right)>0$\label{fig:Lsf1}]{\includegraphics[width=0.3\linewidth]{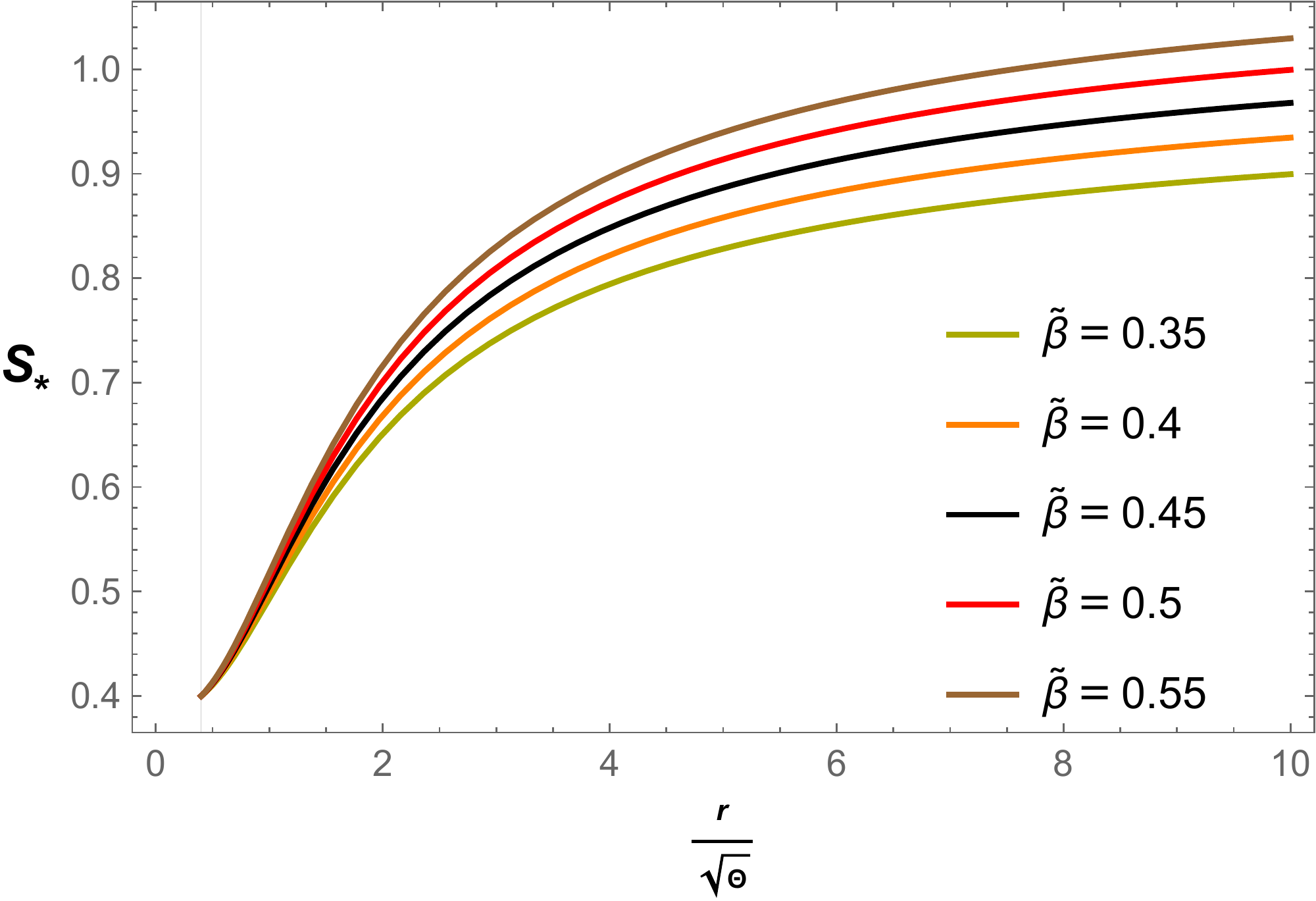}}
    	    \subfloat[$S_*-\frac{r}{\sqrt{\Theta}}<0$\label{fig:Lsf2a}]{\includegraphics[width=0.31\linewidth]{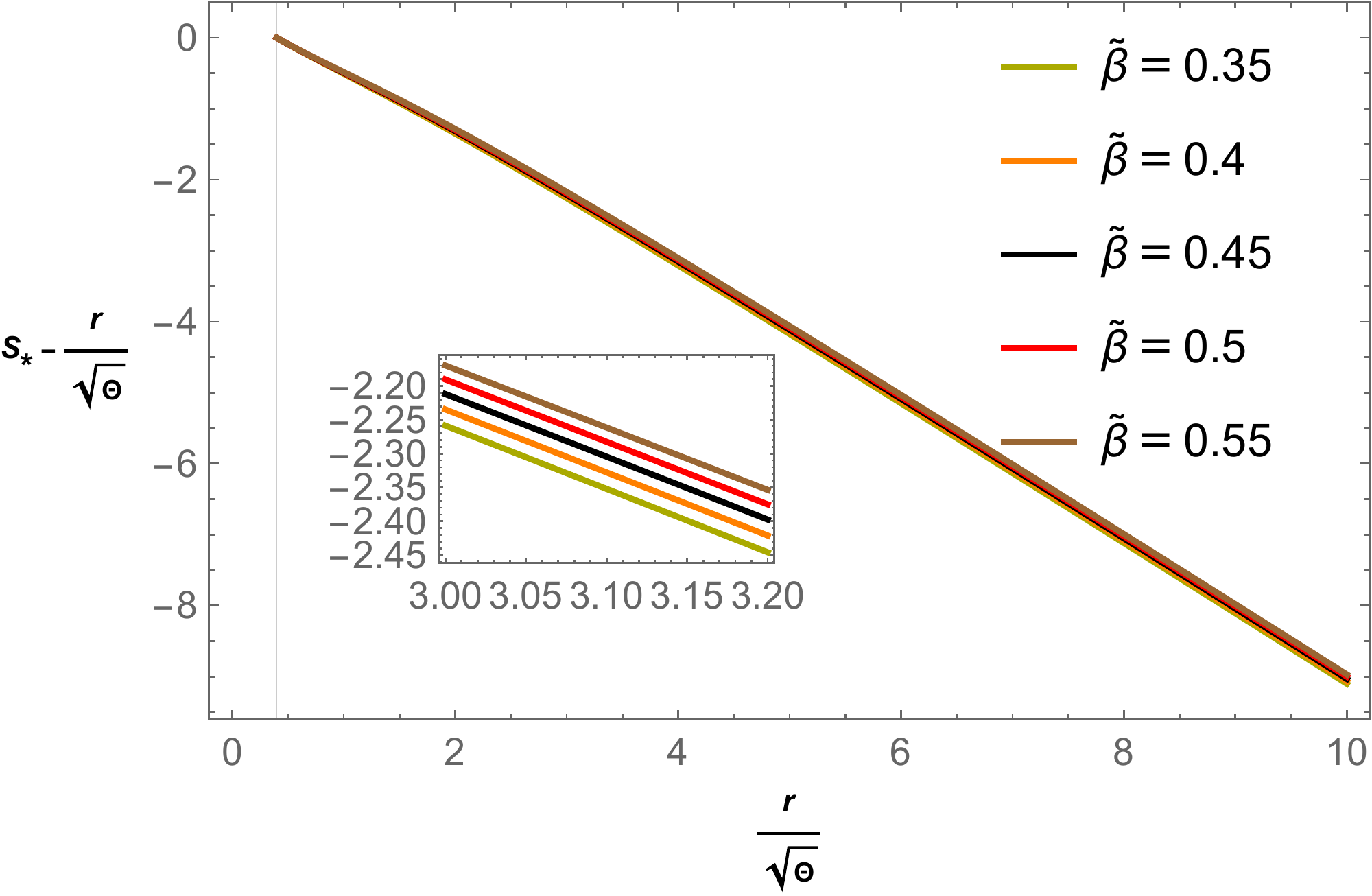}}
    	    \subfloat[$\dot{S}_*\left(\frac{r}{\sqrt{\Theta}}\right)<1$\label{fig:Lsf2b}]{\includegraphics[width=0.3\linewidth]{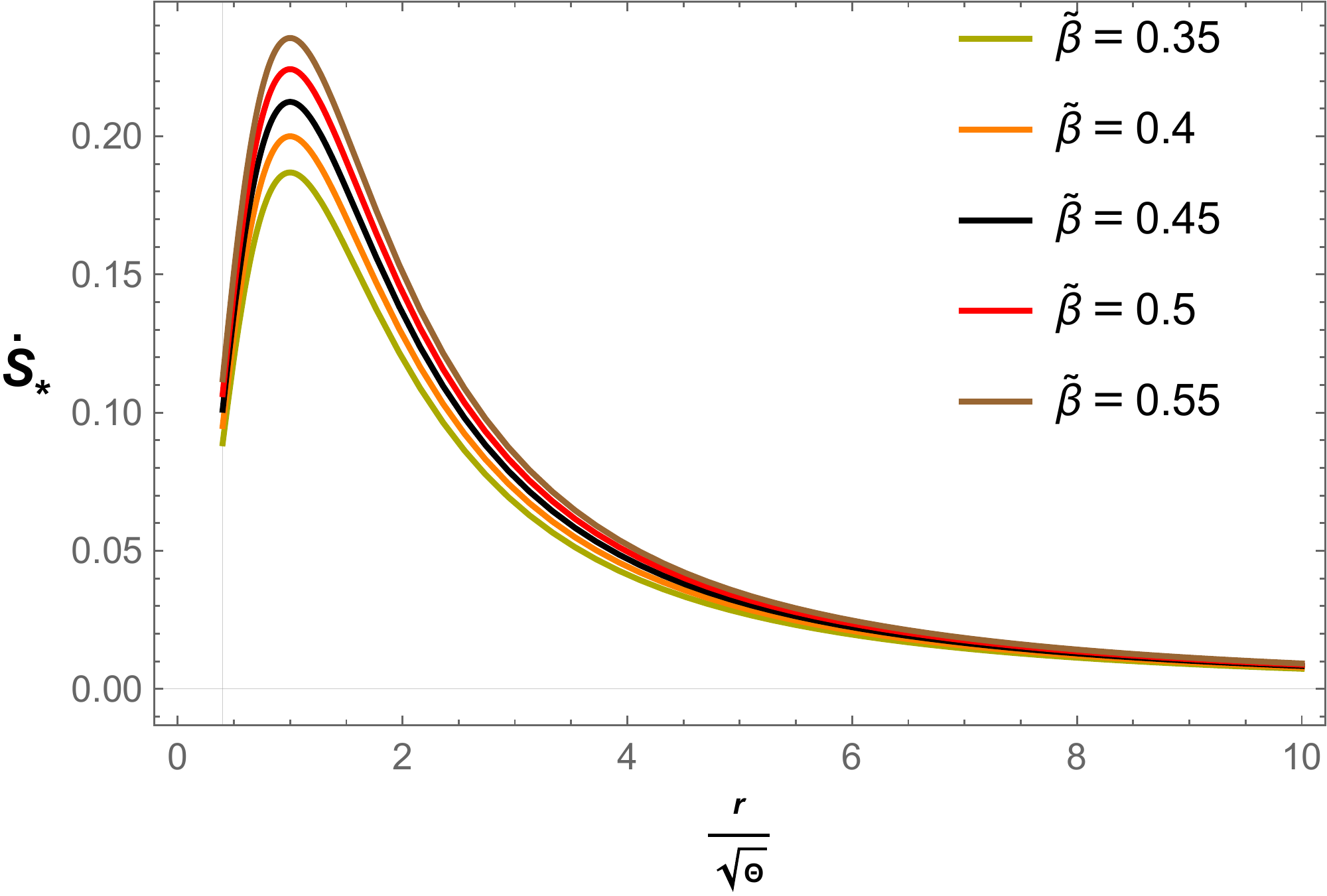}}\\
            \subfloat[$\frac{S_*\left(\frac{r}{\sqrt{\Theta}}\right)-\frac{r}{\sqrt{\Theta}} \dot{S}_*\left(\frac{r}{\sqrt{\Theta}}\right)}{S_*\left(\frac{r}{\sqrt{\Theta}}\right)^2}>0$\label{fig:Lsf3}]{\includegraphics[width=0.3\linewidth]{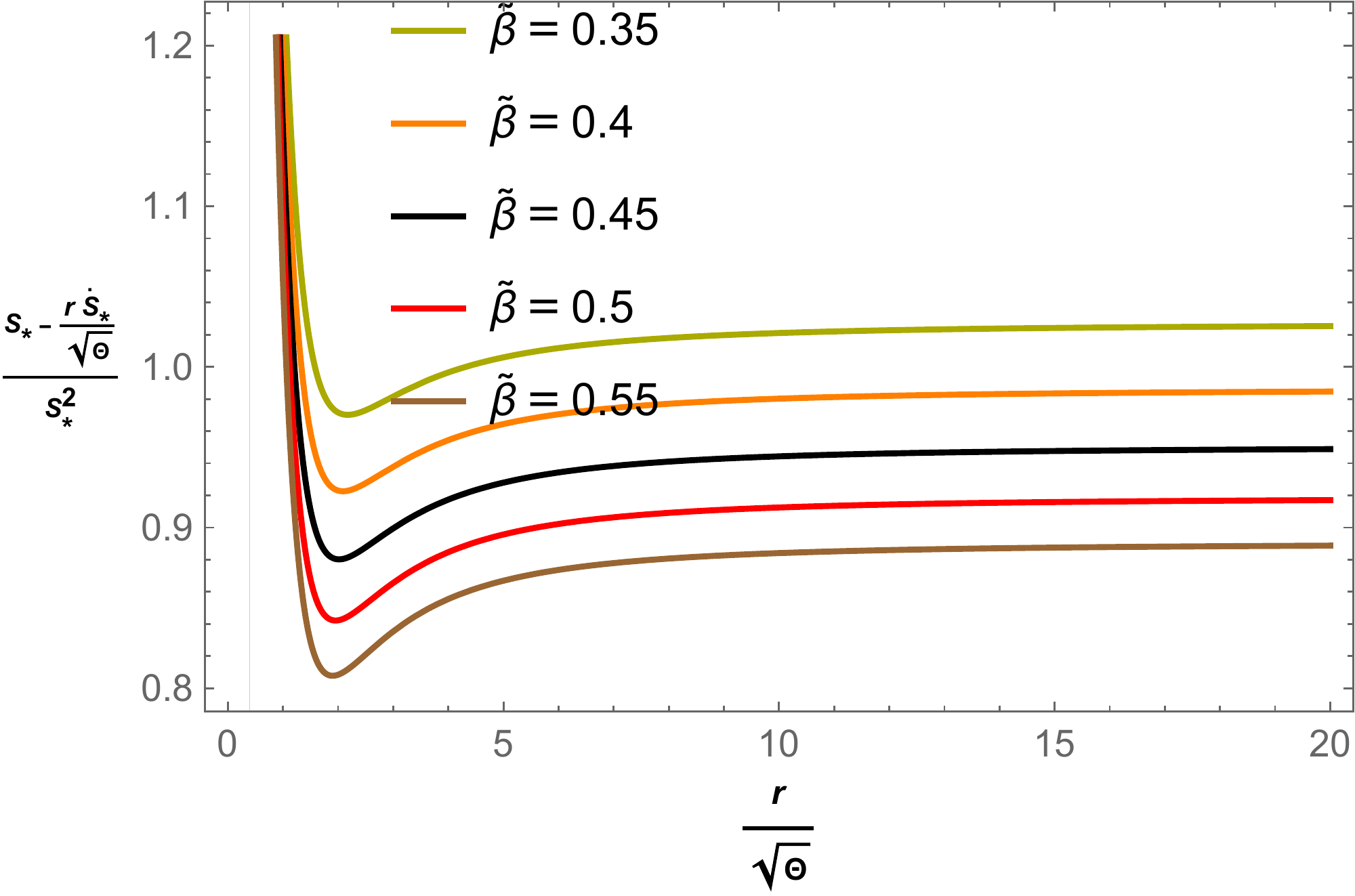}}
            \subfloat[$\frac{S_*\left(\frac{r}{\sqrt{\Theta}}\right)}{\frac{r}{\sqrt{\Theta}}}\to0 $ as $\frac{r}{\sqrt{\Theta}}\to \infty$\label{fig:Lsf4}]{\includegraphics[width=0.3\linewidth]{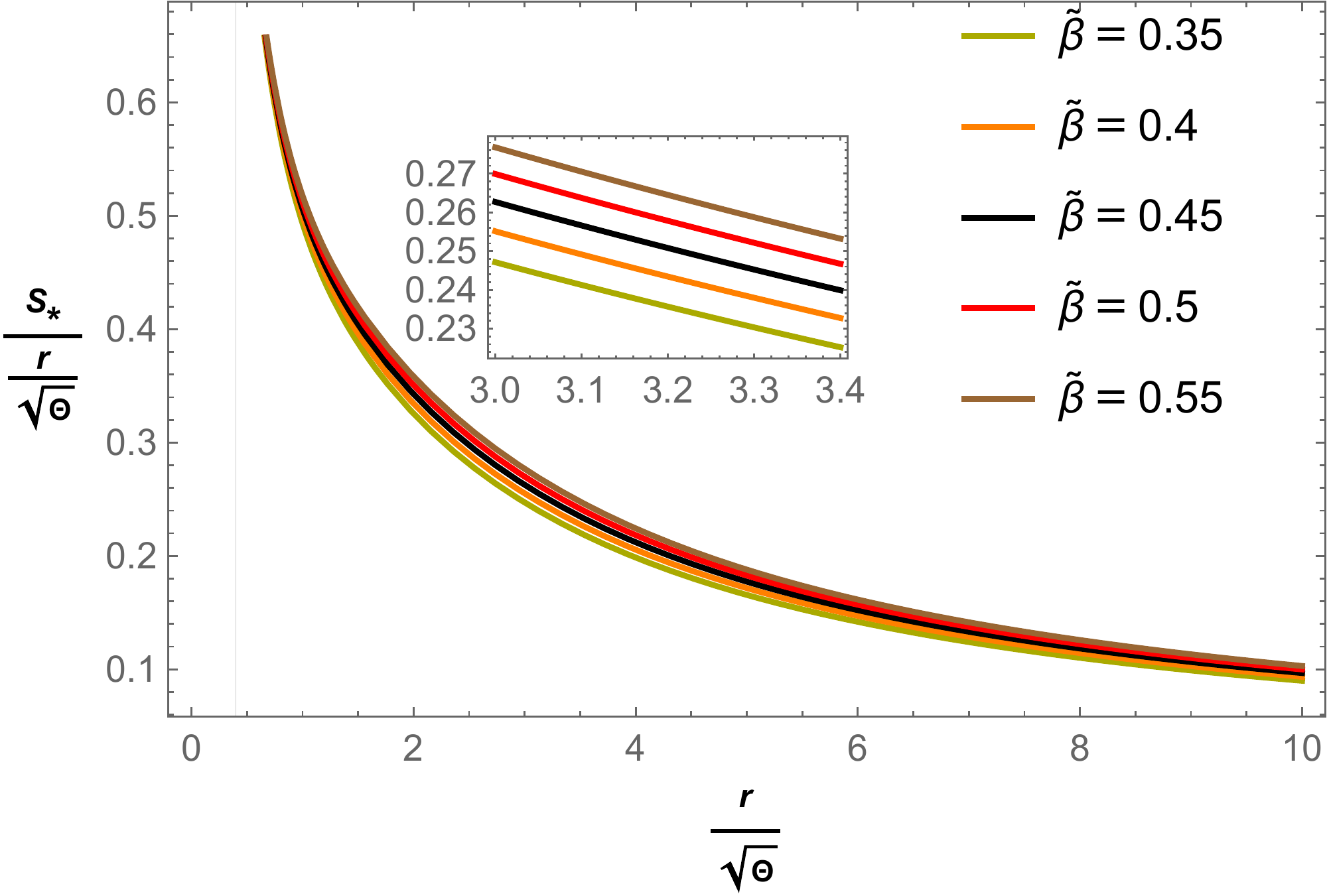}}
            \caption{LNC: Behavior of $S_*$ versus $r/\sqrt{\Theta}$ for different values of $\Tilde{\beta}$ with $\Tilde{M}=3.4$, $\alpha=1.2$, $K_2=2$ and $\frac{r_0}{\sqrt{\Theta}}=0.4$. }
            \label{fig:Lsf}
        \end{figure*}

\begin{figure*}[!]
	    \centering
	    \subfloat[Energy density $\rho$\label{fig:Lrho}]{\includegraphics[width=0.3\linewidth]{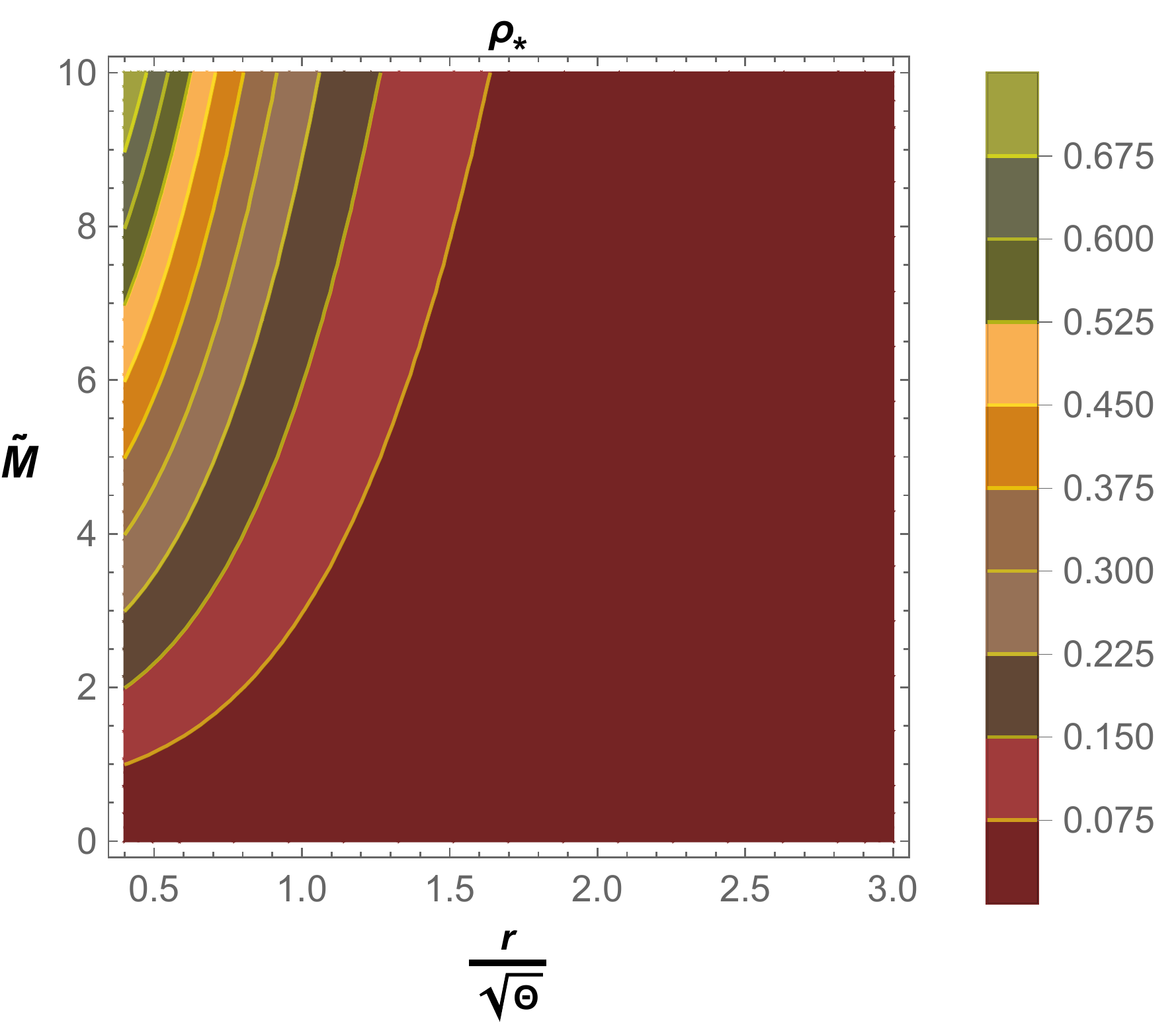}}
	    \subfloat[NEC $\rho+p_r$\label{fig:Le1}]{\includegraphics[width=0.3\linewidth]{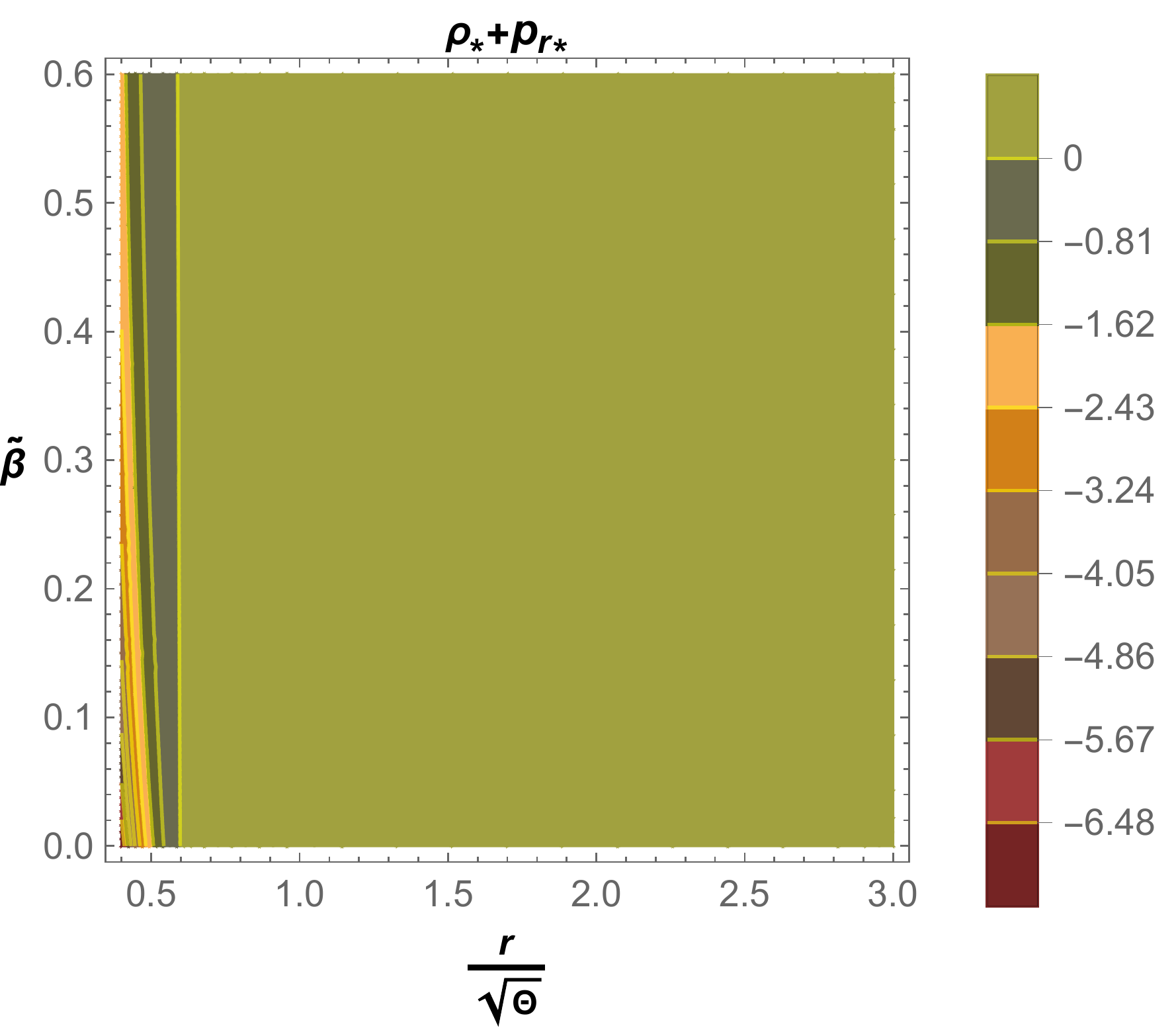}}
	    \subfloat[NEC $\rho+p_\tau$\label{fig:Le2}]{\includegraphics[width=0.3\linewidth]{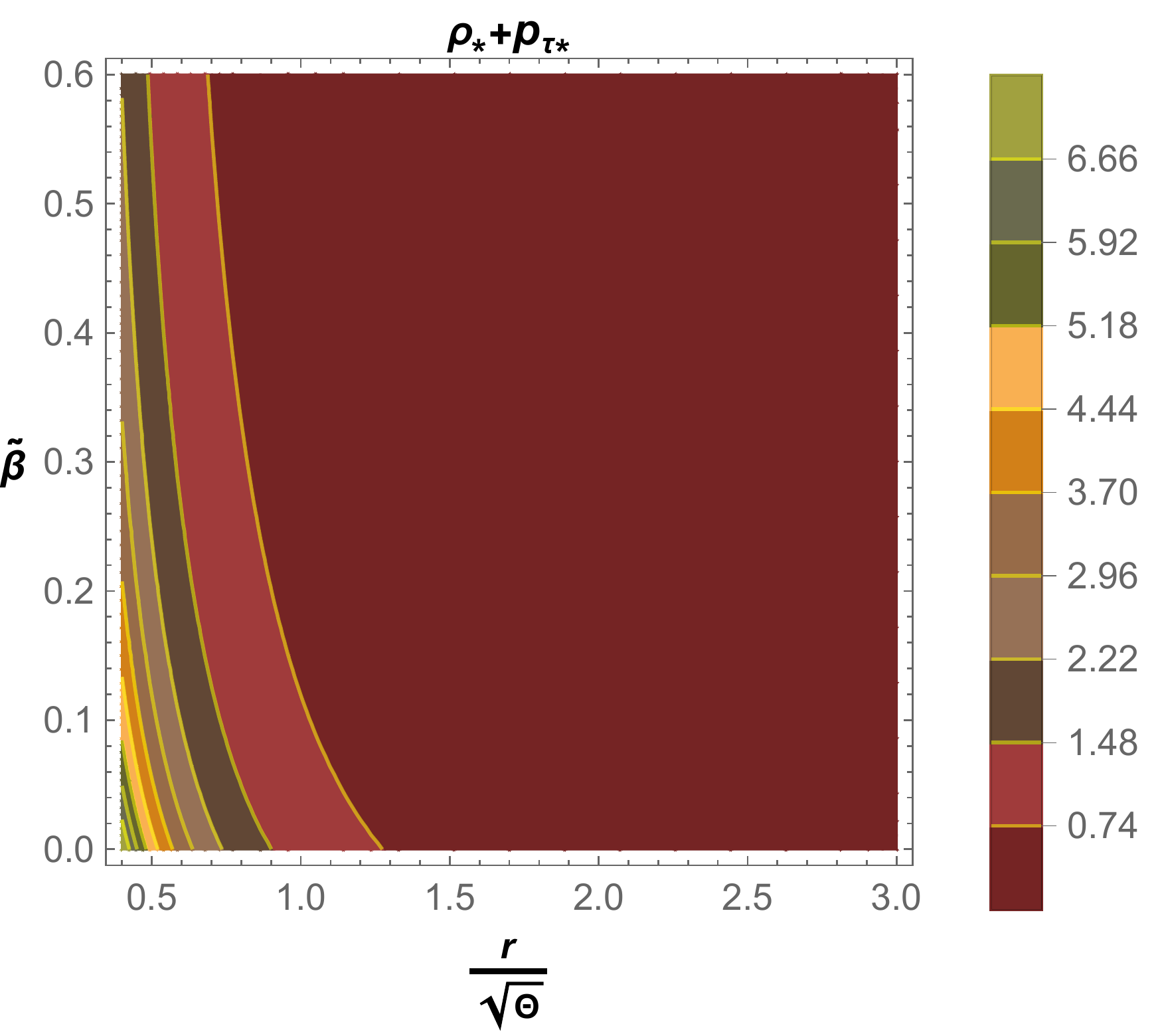}}\\
	    \subfloat[DEC $\rho-|p_r|$\label{fig:Le3}]{\includegraphics[width=0.3\linewidth]{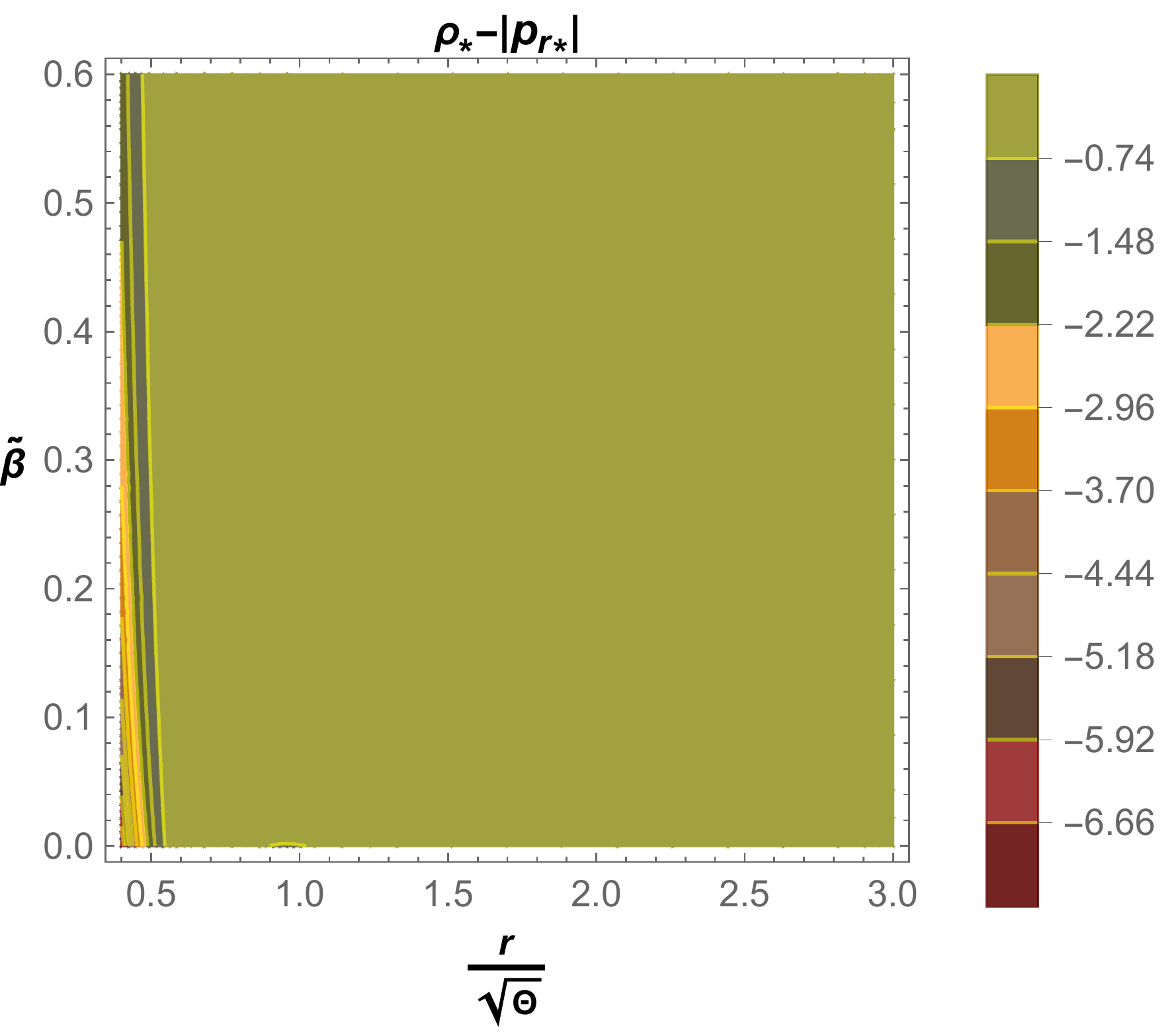}}
	    \subfloat[DEC $\rho-|p_\tau|$\label{fig:Le4}]{\includegraphics[width=0.3\linewidth]{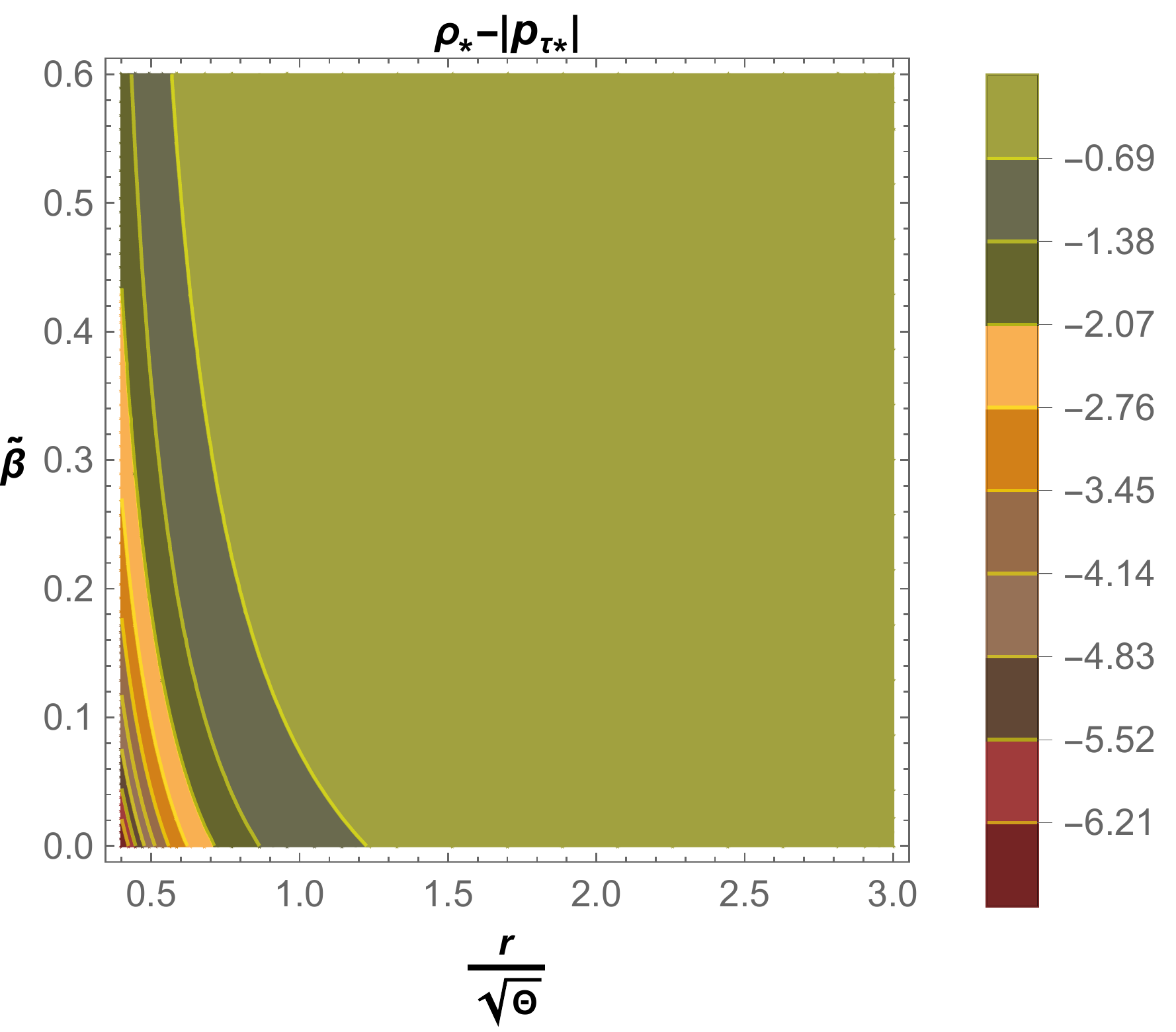}}
	    \subfloat[SEC $\rho+p_r+2p_\tau$\label{fig:Le5}]{\includegraphics[width=0.3\linewidth]{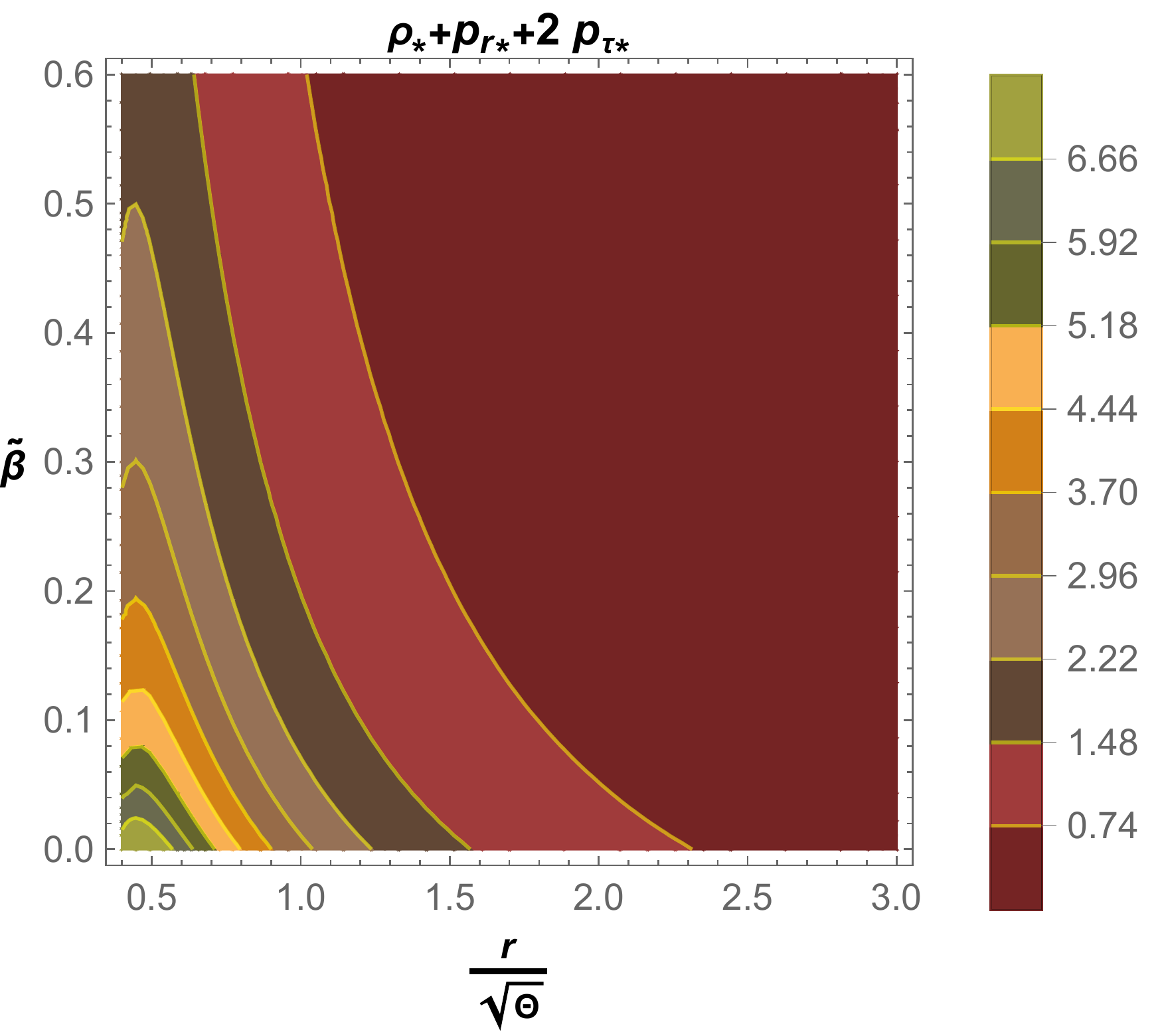}}
	    \caption{LNC: profile of (a) energy density $\rho_*$ varying w.r.t $r/\sqrt{\Theta}$ and $\Tilde{M}$, (b)-(f) different energy conditions varying w.r.t $r/\sqrt{\Theta}$ and model parameter $\Tilde{\beta}$ with $\Tilde{M}=3.4$, $\alpha=1.2$, $K_2=2$ and $\frac{r_0}{\sqrt{\Theta}}=0.4$.}
	    \label{fig:Lec}
	\end{figure*}

\subsection{Lorentzian energy density}
 In the present subsection, our main attention is towards the Lorentzian noncommutative geometry (LNC). By substituting the energy density of LNC \eqref{lorentz} into \eqref{eq:ode}, we get 

 \begin{widetext}
     \begin{equation}
        -\frac{\alpha   \frac{r_0^2}{\Theta} \left[\frac{r}{\sqrt{\Theta }}\dot{H}_{*}\left(\frac{r}{\sqrt{\Theta }}\right)+H_{*}\left(\frac{r}{\sqrt{\Theta }}\right)-K_2\right]}{\frac{r^2}{\Theta}  \left[3 \Tilde{\beta}  \frac{r_0}{\sqrt{\Theta }} \dot{H}_{*}\left(\frac{r_0}{\sqrt{\Theta }}\right)+6 \Tilde{\beta}  H_{*}\left(\frac{r_0}{\sqrt{\Theta }}\right)+K_2 \left(\frac{r_0^2}{\Theta} -2 \Tilde{\beta} \right)\right]}=\frac{ \Tilde{M}}{\pi ^2 \left(1 +\frac{r^2}{\Theta}\right)^2}.
    \end{equation}
 \end{widetext}

    Solving the above equation by imposing the throat condition on the shape function, we can derive the following relation:

    \begin{widetext}
         \begin{equation}
        \begin{split}
            H_*\left(\frac{r}{\sqrt{\Theta}}\right)=\frac{1}{2 \pi ^2 \alpha  \frac{r}{\sqrt{\Theta}} \frac{r_0^2}{\Theta}}&\left[\frac{(\frac{r}{\sqrt{\Theta}}-\frac{r_0}{\sqrt{\Theta}}) }{\left(\frac{r^2}{\Theta}+1\right) \left(\frac{r_0^2}{\Theta}+1\right)}\left\{\Tilde{\beta} \Tilde{M} \left(\frac{r r_0}{\Theta}-1\right) \left(2 K_2-3 n \frac{r_0}{\sqrt{\Theta}}\right)+K_2 \frac{r_0^2}{\Theta} \left(-\Tilde{M} \frac{r r_0}{\Theta}+\Tilde{M}\right.\right.\right.\\&\left.\left.\left.+2 \pi ^2 \alpha  \left(\frac{r^2}{\Theta}+1\right) \left(\frac{r_0^2}{\Theta}+1\right)\right)\right\}+\Tilde{M} \left(\tan ^{-1}\left(\frac{r_0}{\sqrt{\Theta}}\right)-\tan ^{-1}\left(\frac{r}{\sqrt{\Theta}}\right)\right)\right.\\&\left. \times\left(K_2 \left(\frac{r_0^2}{\Theta}-2 \Tilde{\beta}\right)+3 \Tilde{\beta} n \frac{r_0}{\sqrt{\Theta}}\right)\right],
        \end{split}
    \end{equation}
    \end{widetext}

    where
    \begin{equation}
        n=\frac{K_2 \left(\pi ^2 \alpha  \left(\frac{r_0^2}{\Theta}+1\right)^2-\Tilde{M} \left(\frac{r_0^2}{\Theta}-2 \Tilde{\beta}\right)\right)}{3 \Tilde{\beta} \Tilde{M} \frac{r_0}{\sqrt{\Theta}}+\pi ^2 \alpha  \frac{r_0}{\sqrt{\Theta}} \left(\frac{r_0^2}{\Theta}+1\right)^2}.
    \end{equation}

    Therefore, the resulting shape function can be expressed as follows:

\begin{widetext}
      \begin{equation}
        \begin{split}
            S_*\left(\frac{r}{\sqrt{\Theta}}\right)=&\frac{1}{2 \left(\frac{r^2}{\Theta}+1\right) \left(3 \Tilde{\beta} \Tilde{M} \frac{r_0^2}{\Theta}+\pi ^2 \alpha  \left(\frac{r_0^3}{{\Theta}^{\frac{3}{2}}}+\frac{r_0}{\sqrt{\Theta}}\right)^2\right)} \left[\left(\frac{r_0^2}{\Theta}+1\right) \frac{r_0^2}{\Theta} \left\{\Tilde{M} \left(\frac{r}{\sqrt{\Theta}}-\frac{r_0}{\sqrt{\Theta}}\right) \left(\frac{r}{\sqrt{\Theta}} \frac{r_0}{\sqrt{\Theta}}-1\right)\right.\right.\\&\left.\left.+2 \pi ^2 \alpha  \left(\frac{r^2}{\Theta}+1\right) \frac{r_0}{\sqrt{\Theta}} \left(\frac{r_0^2}{\Theta}+1\right)\right\}+\Tilde{M}\left(\frac{r_0^2}{\Theta}+1\right)^2 \left(\Tilde{\beta}+\frac{r_0^2}{\Theta}\right) \left(\tan ^{-1}\left(\frac{r}{\sqrt{\Theta}}\right)-\tan ^{-1}\left(\frac{r_0}{\sqrt{\Theta}}\right)\right)\right.\\&\left.\times \left(\frac{r^2}{\Theta}+1\right) +\Tilde{\beta} \Tilde{M} \left(\frac{r^2}{\Theta} \left(7 \frac{r_0^3}{{\Theta}^{\frac{3}{2}}}+\frac{r_0}{\sqrt{\Theta}}\right)-\frac{r}{\sqrt{\Theta}} \left(\frac{r_0^2}{\Theta}+1\right)^2+7 \frac{r_0^3}{{\Theta}^{\frac{3}{2}}}+\frac{r_0}{\sqrt{\Theta}}\right)\right].
        \end{split}
    \end{equation}
\end{widetext}

    We shall now delve into the characteristics and implications of several aspects pertaining to the obtained shape function. Firstly, $S_*\left(\frac{r}{\sqrt{\Theta}}\right)$ exhibits the property of monotonically increasing function [\figureautorefname~\ref{fig:Lsf1}] and remains consistently lower than the identity function of $\frac{r}{\sqrt{\Theta}}$. Moreover, the imposition of the throat condition directly results in the intersection of the $S_*\left(\frac{r}{\sqrt{\Theta}}\right)-\frac{r}{\sqrt{\Theta}}$ curve with the $\frac{r}{\sqrt{\Theta}}$ axis at the throat point [see \figureautorefname~\ref{fig:Lsf2a}].
    To ensure that the derivative of the shape function remains below 1 at the throat radius, the following constraining relation must be satisfied:

    \begin{equation}
       \Tilde{M} \left(\frac{r_0^2}{\Theta}-2 \Tilde{\beta}\right)< \pi ^2 \alpha  \left(\frac{r_0^2}{\Theta}+1\right)^2.
    \end{equation}

    \figureautorefname~\ref{fig:Lsf2b} and \figureautorefname~\ref{fig:Lsf3} allow us to interpret and confirm that the required condition for the flaring-out behavior satisfied at the throat and beyond ($r>r_0$). As a result, there is a violation of effective NEC at the throat, which is imperative for the existence of a traversable wormhole. Furthermore, as a consequence of $S_*\left(\frac{r}{\sqrt{\Theta}}\right)-\frac{r}{\sqrt{\Theta}}<1$, we have $\frac{S_*\left(\frac{r}{\sqrt{\Theta}}\right)}{\frac{r}{\sqrt{\Theta}}}<1$. For the shape function in hand we have, $\underset{\frac{r}{\sqrt{\Theta}}\to\infty}{\lim}\frac{S_*\left(\frac{r}{\sqrt{\Theta}}\right)}{\frac{r}{\sqrt{\Theta}}}=0$, implying the fulfilment of asymptotic flatness condition [\figureautorefname~\ref{fig:Lsf4}].
    
    For LNC, the pressure elements can be expressed as, 
    
\begin{widetext}
    \begin{gather}
    \begin{split}
         p_{r*}=\frac{1}{2 \pi ^2 \frac{r^3}{\Theta^{\frac{3}{2}}}}&\left[\frac{1}{\left(\frac{r_0^2}{\Theta}+1\right)^2}\left\{\frac{3 \Tilde{M} \left(-7 \left(\frac{r^2}{\Theta}+1\right) \frac{r_0^3}{{\Theta}^{\frac{3}{2}}}+2 \frac{r}{\sqrt{\Theta}} \left(2 \frac{r^2}{\Theta}+3\right) \frac{r_0^2}{\Theta}-\left(\frac{r^2}{\Theta}+1\right) \frac{r_0}{\sqrt{\Theta}}+\frac{r}{\sqrt{\Theta}} \frac{r_0}{\sqrt{\Theta}}^4+\frac{r}{\sqrt{\Theta}}\right)}{\frac{r^2}{\Theta}+1}\right.\right.\\&\left.\left.+\frac{2 \left(2 \frac{r}{\sqrt{\Theta}}-3 \frac{r_0}{\sqrt{\Theta}}\right) \left(\pi ^2 \alpha  \left(\frac{r_0^3}{{\Theta}^{\frac{3}{2}}}+\frac{r_0}{\sqrt{\Theta}}\right)^2-3 \Tilde{M} \frac{r_0}{\sqrt{\Theta}}^4\right)}{\Tilde{\beta}+\frac{r_0^2}{\Theta}}\right\}-3 \Tilde{M} \tan ^{-1}\left(\frac{r}{\sqrt{\Theta}}\right)+3 \Tilde{M} \tan ^{-1}\left(\frac{r_0}{\sqrt{\Theta}}\right)\right],
    \end{split}
     \\
      p_{\tau*}=\frac{1}{\frac{r^2}{\Theta} \left(\Tilde{\beta}+\frac{r_0^2}{\Theta}\right)}\left[\frac{\Tilde{M} \left(3 \Tilde{\beta} \frac{r^4}{\Theta^2} \frac{r_0^2}{\Theta}-\frac{r^2}{\Theta} \left(\Tilde{\beta}+\frac{r_0}{\sqrt{\Theta}}^6+(\Tilde{\beta}+2) \frac{r_0}{\sqrt{\Theta}}^4+(1-4 \Tilde{\beta}) \frac{r_0^2}{\Theta}\right)+3 \Tilde{\beta} \frac{r_0^2}{\Theta}\right)}{\pi ^2 \left(\frac{r^2}{\Theta}+1\right)^2 \left(\frac{r_0^2}{\Theta}+1\right)^2}+\alpha  \frac{r_0^2}{\Theta}\right].
  \end{gather}

\end{widetext}
    
  The behavior of the energy conditions, along with the accompanying energy density profile for LNC, is effectively demonstrated in \figureautorefname~\ref{fig:Lec}. It reveals that the radial NEC and both DECs are violated in this context. However, the tangential NEC and the SEC are obeyed.
  
   \begin{figure}[h!]
            \centering
            \subfloat[Characteristics of $S_*$ with GNC\label{fig:TGSF}]{\includegraphics[width=0.7\linewidth]{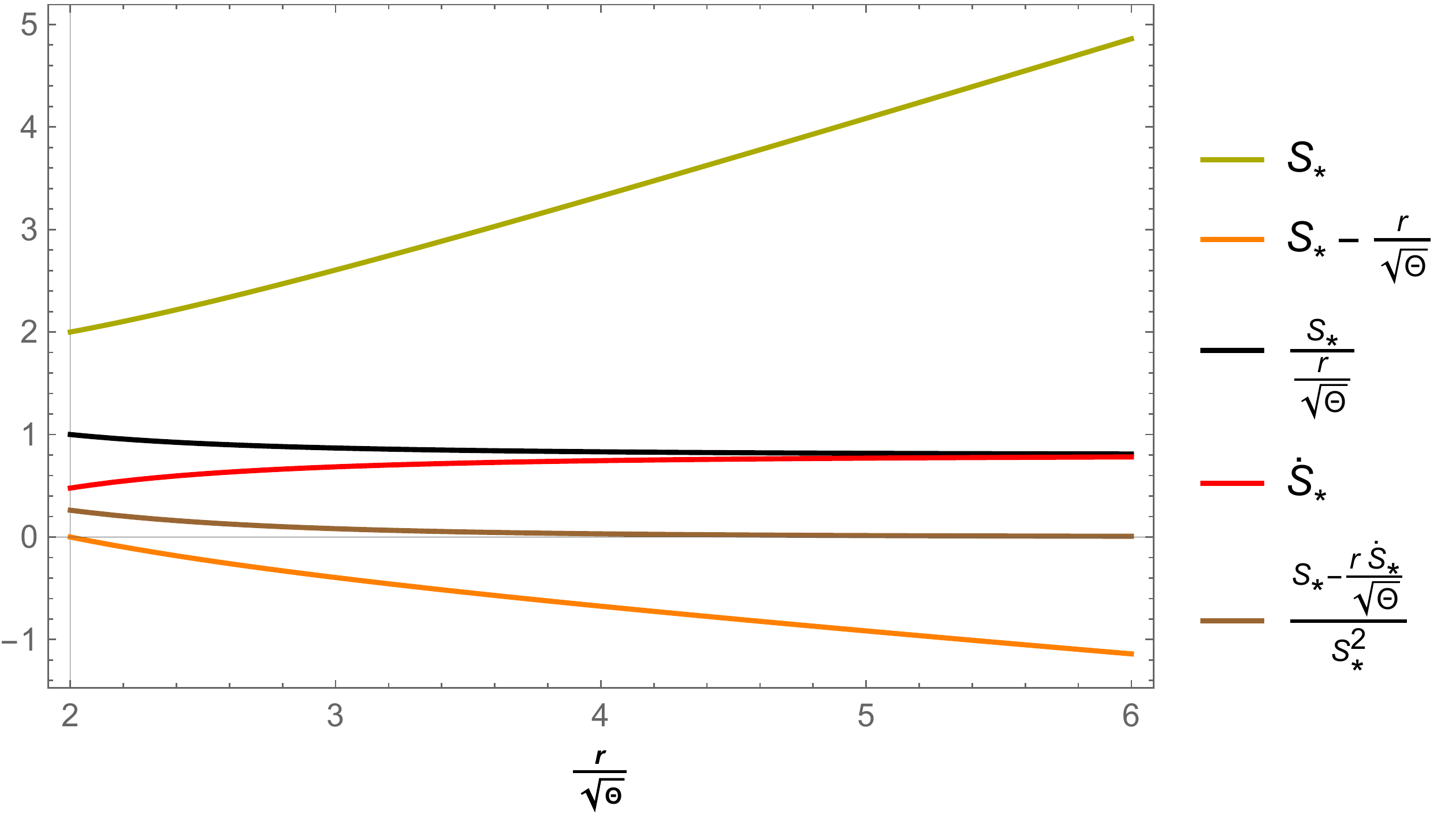}}\\
    	    \subfloat[Characteristics of $S_*$ with LNC\label{fig:TLSF}]{\includegraphics[width=0.7\linewidth]{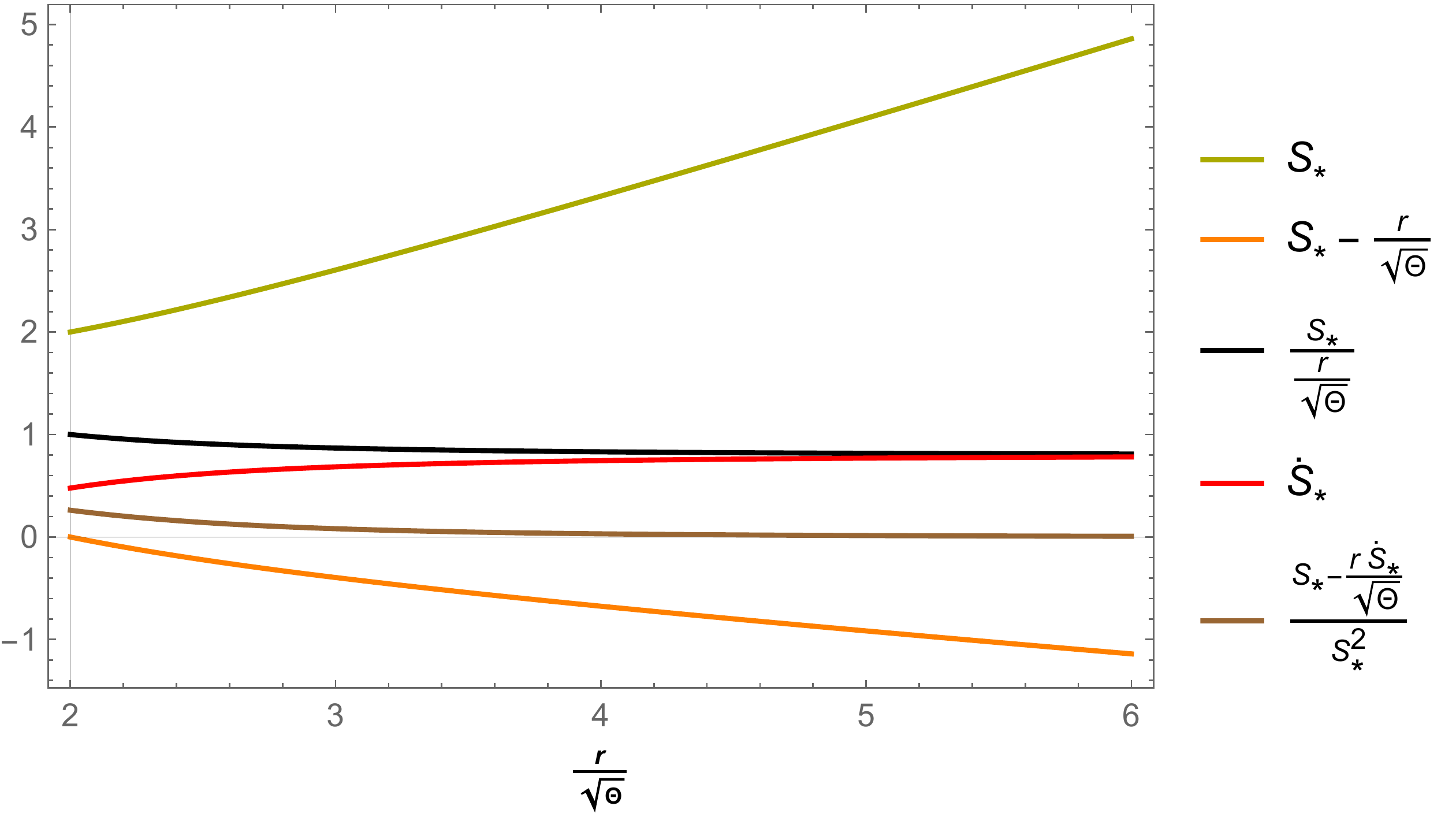}}
            \caption{Traceless fluid: Behavior of $S_*$ versus $r/\sqrt{\Theta}$ for $\Tilde{\beta}=1.2$, $\Tilde{M}=4.5$, $\alpha=1.4$, $K_2=7$ and $\frac{r_0}{\sqrt{\Theta}}=2$.}
            \label{fig:Tsf}
        \end{figure}

\section{Wormhole Solutions with Traceless Fluid}\label{IV}

In this section, we shall focus on the matter with a traceless fluid \cite{ec1,traceless}. It is characterized by a specific form of EoS $\rho-p_r-2p_\tau=0$ or equivalently, $\rho_*-p_{r*}-2p_{\tau*}=0$. Now, by substituting the pressure expressions \eqref{eq:pr} and \eqref{eq:pt}, along with the noncommutative energy densities \eqref{gauss} and \eqref{lorentz}, into the equations, we obtain the following set of differential equations for the traceless fluid:

\begin{widetext}
    
\begin{gather}
     \label{ode1}\frac{8 \pi ^{3/2} \alpha  \frac{r_0^2}{\Theta} \left[2 \frac{r}{\sqrt{\Theta}} H_*'\left(\frac{r}{\sqrt{\Theta}}\right)+5 H_*\left(\frac{r}{\sqrt{\Theta}}\right)-K_2\right]}{\frac{r}{\sqrt{\Theta}} \left[K_2\left(\frac{r_0^2}{\Theta}-2 \Tilde{\beta}\right)+3 \Tilde{\beta} H_*'\left(\frac{r_0}{\sqrt{\Theta}}\right) \frac{r_0}{\sqrt{\Theta}}\right]}=\Tilde{M} e^{-\frac{r^2}{4\Theta}} \frac{r}{\sqrt{\Theta}},~ \text{and}\\
     \label{ode2}\frac{2 \alpha  \frac{r_0^2}{\Theta} \left[\frac{r}{\sqrt{\Theta}} H_*'\left(\frac{r}{\sqrt{\Theta}}\right)+H_*\left(\frac{r}{\sqrt{\Theta}}\right)\right]}{\frac{r^2}{\Theta} \left[K_2 \left(\frac{r_0^2}{\Theta}-2 \Tilde{\beta}\right)+3 \Tilde{\beta} n \frac{r_0}{\sqrt{\Theta}}\right]}=\frac{\alpha  \frac{r_0^2}{\Theta} (K_2-3 H_*\left(\frac{r}{\sqrt{\Theta}}\right)}{\frac{r^2}{\Theta} \left(K_2 \left(\frac{r_0^2}{\Theta}-2 \Tilde{\beta}\right)+3 \Tilde{\beta} H_*'\left(\frac{r_0}{\sqrt{\Theta}}\right) \frac{r_0}{\sqrt{\Theta}}\right)}+\frac{\Tilde{M}}{\pi ^2 \left(\frac{r^2}{\Theta}+1\right)^2}.
\end{gather}
\end{widetext}

Upon solving equations \eqref{ode1} and \eqref{ode2} for $H_*$, and using the relation \eqref{eq:main} along with the application of the throat condition, we obtain the exact adimensional shape functions as,

\begin{widetext}
    \begin{equation}
    \begin{split}
       S_*\left(\frac{r}{\sqrt{\Theta}}\right)=\frac{4 \left(4 \left(\frac{r}{\sqrt{\Theta}}\right)^{5/2}+\left(\frac{r_0}{\sqrt{\Theta}}\right)^{5/2}\right)-\frac{5 \Tilde{M} e^{\frac{r_0^2}{4\Theta}} \left(2 \frac{r_0^2}{\Theta}-\Tilde{\beta}\right) \left(\left(\frac{r_0}{\sqrt{\Theta}}\right)^{9/2} E_{-\frac{5}{4}}\left(\frac{r_0^2}{4\Theta}\right)-\left(\frac{r}{\sqrt{\Theta}}\right)^{9/2} E_{-\frac{5}{4}}\left(\frac{r^2}{4\Theta}\right)\right)}{\frac{r_0^2}{\Theta} \left(16 \pi ^{3/2} \alpha  e^{\frac{r_0^2}{4\Theta}}-3 \Tilde{\beta} \Tilde{M}\right)}}{20 \left(\frac{r}{\sqrt{\Theta}}\right)^{3/2}},
    \end{split}
\end{equation}

    \begin{equation}
    \begin{split}
         S_*\left(\frac{r}{\sqrt{\Theta}}\right)=&\frac{1}{2 \left(\frac{r}{\sqrt{\Theta}}\right)^{3/2}}\left[-\frac{5 \Tilde{M} \left(\frac{r_0^2}{\Theta}+1\right)^2 \left(2 \frac{r_0^2}{\Theta}-\Tilde{\beta}\right) }{48 \frac{r_0^2}{\Theta} \left(2 \pi ^2 \alpha  \left(\frac{r_0^2}{\Theta}+1\right)^2-3 \Tilde{\beta} \Tilde{M}\right)}\left\{\frac{24 \sqrt{\frac{r}{\sqrt{\Theta}}}}{\frac{r^2}{\Theta}+1}+3 \sqrt{2} \left(-\log \left(\frac{r}{\sqrt{\Theta}}+\sqrt{2} \sqrt{\frac{r}{\sqrt{\Theta}}}+1\right)\right.\right.\right.\\&\left.\left.\left.+\log \left(\frac{r}{\sqrt{\Theta}}-\sqrt{2} \sqrt{\frac{r}{\sqrt{\Theta}}}+1\right)-2 \tan ^{-1}\left(\sqrt{2} \sqrt{\frac{r}{\sqrt{\Theta}}}+1\right)+2 \tan ^{-1}\left(1-\sqrt{2} \sqrt{\frac{r}{\sqrt{\Theta}}}\right)\right)\right\}\right.\\&\left.+\frac{2 \Tilde{M} \left(\frac{r}{\sqrt{\Theta}}\right)^{5/2} \left(\frac{r_0^2}{\Theta}+1\right)^2 \left(2 \frac{r_0^2}{\Theta}-\Tilde{\beta}\right)}{\left(\frac{r}{\sqrt{\Theta}}+1\right) \left(3 \Tilde{\beta} \Tilde{M} \frac{r_0^2}{\Theta}-2 \pi ^2 \alpha  \left(\left(\frac{r_0}{\sqrt{\Theta}}\right)^3+\frac{r}{\sqrt{\Theta}}\right)^2\right)}+\frac{2 \Tilde{M} \left(\frac{r_0^2}{\Theta}+1\right) \sqrt{\frac{r_0}{\sqrt{\Theta}}} \left(2 \frac{r_0^2}{\Theta}-\Tilde{\beta}\right)}{2 \pi ^2 \alpha  \left(\frac{r_0^2}{\Theta}+1\right)^2-3 \Tilde{\beta} \Tilde{M}}\right.\\&\left.+\frac{5 \Tilde{M} \left(\frac{r_0^2}{\Theta}+1\right)^2 \left(2 \frac{r_0^2}{\Theta}-\Tilde{\beta}\right)}{48 \frac{r_0^2}{\Theta} \left(2 \pi ^2 \alpha  \left(\frac{r_0^2}{\Theta}+1\right)^2-3 \Tilde{\beta} \Tilde{M}\right)}\left\{\frac{24 \sqrt{\frac{r_0}{\sqrt{\Theta}}}}{\frac{r_0^2}{\Theta}+1}+3 \sqrt{2} \left(-\log \left(\frac{r_0}{\sqrt{\Theta}}+\sqrt{2} \sqrt{\frac{r_0}{\sqrt{\Theta}}}+1\right)\right.\right.\right.\\&\left.\left.\left.+\log \left(\frac{r_0}{\sqrt{\Theta}}-\sqrt{2} \sqrt{\frac{r_0}{\sqrt{\Theta}}}+1\right)-2 \tan ^{-1}\left(\sqrt{2} \sqrt{\frac{r_0}{\sqrt{\Theta}}}+1\right)+2 \tan ^{-1}\left(1-\sqrt{2} \sqrt{\frac{r_0}{\sqrt{\Theta}}}\right)\right)\right\}\right.\\&\left.+\frac{8 \left(\frac{r}{\sqrt{\Theta}}\right)^{5/2}}{5}+\frac{2 \left(\frac{r_0}{\sqrt{\Theta}}\right)^{5/2}}{5}\right].
    \end{split}
    \end{equation}
\end{widetext}

where $E_n(z)=\int_1^{\infty}e^{-zt}/t^n$ is an exponential integral function. From these expressions, the derivative at the throat can be obtained as
\begin{gather}
    \label{flaring1}\dot{S}_*\left(\frac{r_0}{\sqrt{\Theta}}\right)=\frac{1}{2} \left(\frac{\Tilde{M} \left(\Tilde{\beta}-2 \frac{r_0^2}{\Theta}\right)}{16 \pi ^{3/2} \alpha  e^{\frac{r_0^2}{4\Theta}}-3 \Tilde{\beta} \Tilde{M}}+1\right),~\text{and}\\
   \label{flaring2} \dot{S}_*\left(\frac{r_0}{\sqrt{\Theta}}\right)=\frac{\pi ^2 \alpha  \left(\frac{r_0^2}{\Theta}+1\right)^2-\Tilde{M} \left(\Tilde{\beta}+\frac{r_0^2}{\Theta}\right)}{2 \pi ^2 \alpha  \left(\frac{r_0^2}{\Theta}+1\right)^2-3 \Tilde{\beta} \Tilde{M}}.
\end{gather}

\begin{figure}[!]
            \centering
            \subfloat[GNC\label{fig:TGEC}]{\includegraphics[width=0.9\linewidth]{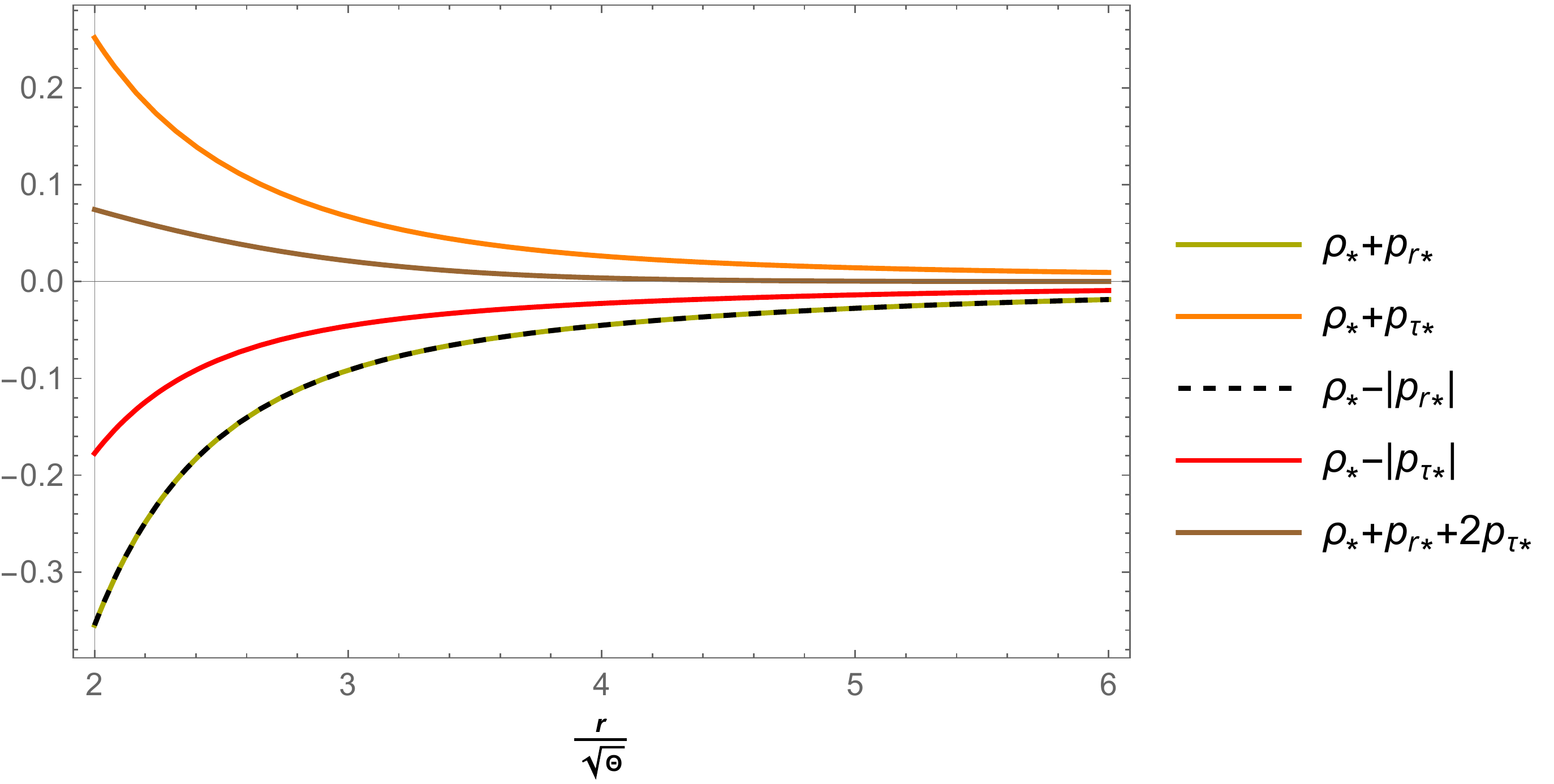}}\\
    	    \subfloat[LNC\label{fig:TLEC}]{\includegraphics[width=0.9\linewidth]{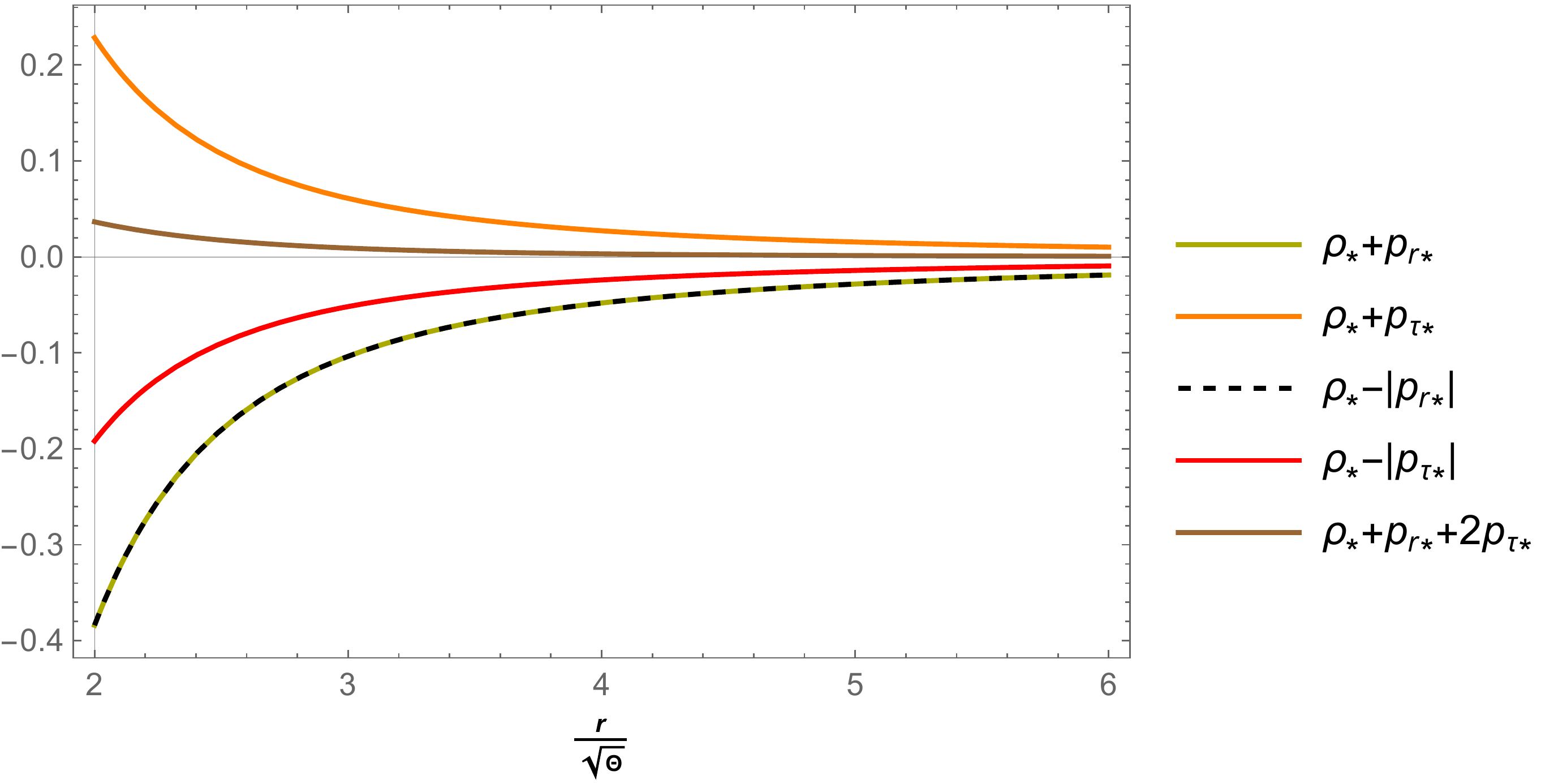}}
            \caption{Traceless fluid: Different energy conditions varying w.r.t $r/\sqrt{\Theta}$ with $\Tilde{\beta}=1.2$, $\Tilde{M}=4.5$, $\alpha=1.4$, $K_2=7$ and $\frac{r_0}{\sqrt{\Theta}}=2$..}
            \label{fig:Tec}
        \end{figure}

To satisfy the flaring-out conditions at the throat, equations \eqref{flaring1} and \eqref{flaring2} should both be less than 1. In order to achieve this, we have chosen the following parameter values: $\Tilde{M} = 4.5$, $\frac{r_0}{\sqrt{\Theta}} = 2$, $\alpha = 1.4$, $\Tilde{\beta} = 1.2$, and $K_2 = 7$. With these values, the flaring-out condition is indeed satisfied. Unfortunately, $\underset{\frac{r}{\sqrt{\Theta}}\to\infty}{\lim}\frac{S_*\left(\frac{r}{\sqrt{\Theta}}\right)}{\frac{r}{\sqrt{\Theta}}}=\frac{4}{5}$ implies the disobeying asymptotic flatness condition of $S_*$. However, it should be noted that the convergence point has an extremely small magnitude. Similar results have been reported in \cite{asymptotic}. The behavior of shape functions is illustrated in \figureautorefname~\ref{fig:Tsf}. Furthermore, we verified the energy conditions. For both wormholes, radial NEC is violated, while tangential NEC is satisfied. Both DECs are violated and SEC is obeyed [see \figureautorefname~\ref{fig:Tec}].


\section{Results and Concluding Remarks}\label{V}

In this study, we have conducted a comprehensive investigation into wormhole solutions within the framework of modified gravity theories incorporating noncommutative backgrounds and conformal symmetries. We have thoroughly examined the viability of traversable wormholes under various scenarios. The incorporation of noncommutative geometries has provided a novel perspective on the manifold structure and its influence on wormhole physics. We have taken into account the discretization of spacetime and the smearing effect associated with noncommutativity, resulting in distributed energy densities instead of localized point-like sources. The chosen Gaussian and Lorentzian distributions have demonstrated the regularization effects of noncommutativity, ensuring finite and asymptotically vanishing physical parameters ($\rho,p_r,p_\tau$) away from the origin. The conformal factor and CVKs have played pivotal roles in elucidating the conformal motion and metric transformations inherent in wormhole solutions. By exploring curvature-matter coupling gravity theories, particularly within the context of $\mathpzc{f}(\mathcal{R},\mathscr{L}_m)$ gravity, we have examined the wormhole solutions and assessed the effects of modified field equations on the geometry and matter content of wormholes. $\mathpzc{f}(\mathcal{R},\mathscr{L}_m)$ gravity is one of the prominent extensions of $\mathpzc{f}(\mathcal{R})$ theory. Both $\mathpzc{f}(\mathcal{R})$ and $\mathpzc{f}(\mathcal{R},\mathscr{L}_m)$ are indeed prominent modified theories of gravity. They can address various cosmological and astrophysical issues through their underlying geometric and theoretical frameworks. However, a point of particular interest in $\mathpzc{f}(\mathcal{R},\mathscr{L}_m)$ theory is its approach to handling the coupling effect between geometry and matter \cite{frlm1,frlm2,frlm3,frlm4,frlm5,frlm6,frlm7,frlm8}. In \cite{comp1,comp2}, similar kinds of approaches are used to study wormhole solutions within the background of $\mathpzc{f}(\mathcal{R},\phi)$ coupling and modified $\mathpzc{f}(\mathcal{R})$ gravity. The salient aspects of the present study are outlined as follows:
\begin{itemize}
    \item We have examined a gravitational theory with a curvature-matter coupling, represented by $\mathpzc{f}(\mathcal{R},\mathscr{L}_m)=\alpha\dfrac{\mathcal{R}}{2}+(1+\beta\mathcal{R}_0)\mathscr{L}_m,$ where $\alpha$ and $\beta$ are model parameters. Here, the modification is mainly focused on the coefficient of the matter Lagrangian. This particular functional form of curvature and matter coupling presents an interesting aspect of our study. Notably, we have adopted a unique approach by treating the throat radius, $r_0$, as a variable rather than a fixed value. However, the ratio $r_0/\sqrt{\Theta}$ is fixed.

    \item The Lagrangian serves to analyze the underlying physical characteristics of spacetime and plays a pivotal role in the examination of the trajectories of test particles. The selection of the matter Lagrangian offers a precise prescription for determining the distribution of matter within spacetime by presenting the associated energy-momentum candidate. In this work, we focused on exploring scenarios where matter density exhibits an anisotropic nature. For this purpose, we opted for the Lagrangian density $\mathscr{L}_m$ as a function of the average pressure $P$, denoted as $\mathscr{L}_m=P$.
    
    \item With GNC and LNC we have derived the corresponding shape functions by imposing throat condition. Further, we examined the influence of model parameters on them. Both the obtained shape functions obeyed all the essential criteria. These are illustrated in \figureautorefname~\ref{fig:Gsf} and \ref{fig:Lsf}. 
    
    \item The determination of parameter values in this study is guided by the constraints imposed to ensure the fulfillment of essential wormhole properties. These constraints serve as guiding principles for selecting appropriate parameter values that align with the desired characteristics and behaviors of wormholes.

    \item Next, we examined traceless wormhole solutions with GNC and LNC and obtained viable shape functions. The derived shape functions fulfill all the necessary conditions, except for asymptotic flatness, which is consistent with the findings of \cite{asymptotic}.
    
    \item Due to CKVs we could not obtain an asymptotically vanishing lapse function. This result is similar to the previous studies \cite{ckv4,ckv5,ckv6,traceless}.
    
    \item For all the wormhole solutions the NEC is found to be violating conforming the requirement of hypothetical fluid. \figureautorefname~\ref{fig:Gec}, \ref{fig:Lec} and \ref{fig:Tec} depicts the energy conditions of the corresponding cases. In \figureautorefname~\ref{fig:Gec} and \ref{fig:Lec}, we have plotted contours that illustrate the variation of the quantity concerning both the adimensional parameter and the model parameter. The scale on the right side of each graph indicates how the model parameter influences the value of the corresponding physical quantities in conjunction with the changes in the adimensional coordinate $(r/\sqrt{\Theta})$.
\end{itemize}

To conclude, in this manuscript, we have explored new wormhole solutions in the context of three pivotal factors. Our choice of coupling theory offered an extended perspective rooted in the $\mathpzc(\mathcal{R})$ theory and has exhibited a coupling effect in the analysis of wormhole solution.  The implementation of noncommutative geometry and CVKs introduced quantization effects through the discretization of spacetime. Thus, our work investigated the plausibility of traversable wormhole solutions within the framework of these modifications.



\section*{Data Availability Statement}
There are no new data associated with this article.

\begin{acknowledgments}
N.S.K. and V.V. acknowledge DST, New Delhi, India, for its financial support for research facilities under DST-FIST-2019. 
\end{acknowledgments}


\end{document}